\documentclass[a4paper,fleqn,usenatbib]{mnras}

\usepackage[T1]{fontenc}
\usepackage{ae,aecompl}

\usepackage{graphicx}	
\usepackage{amsmath}	
\usepackage{amssymb}	
\usepackage{xcolor}
\usepackage{enumitem}

\title[]{Evolution of fractality and rotation in embedded star clusters}

\author[A. Ballone et al.]{Alessandro Ballone$^{1,2,3}$, Michela Mapelli$^{1,2,3}$, Ugo N. Di Carlo$^{2,3,4}$, 
\newauthor Stefano Torniamenti$^{1,3}$, Mario Spera$^{1,3,5,6}$, Sara Rastello$^{1,3}$
\\
$^{1}$Physics and Astronomy Department Galileo Galilei, University of Padova, Vicolo dell'Osservatorio 3, I-35122 Padova, Italy\\
$^{2}$INAF - Osservatorio Astronomico di Padova, Vicolo dell'Osservatorio 5, I-35122 Padova, Italy\\
$^{3}$INFN - Padova, Via Marzolo 8, I--35131 Padova, Italy\\
$^{4}$Dipartimento di Scienza e Alta Tecnologia, University of Insubria, Via Valleggio 11, I-22100 Como, Italy\\
$^{5}$Center for Interdisciplinary Exploration and Research in Astrophysics (CIERA), Evanston, IL 60208, USA\\
$^{6}$Department of Physics \& Astronomy, Northwestern University, Evanston, IL 60208, USA\\
}

\date{Accepted XXX. Received YYY; in original form ZZZ}

\pubyear{2018}

\begin{document}
\label{firstpage}
\pagerange{\pageref{firstpage}--\pageref{lastpage}}
\maketitle

\begin{abstract}
More and more observations indicate that young star clusters could retain imprints of their formation process. In particular, the degree of substructuring and rotation are possibly the direct result of the collapse of the parent molecular cloud from which these systems form. Such properties can, in principle, be washed-out, but they are also expected to have an impact on the relaxation of these systems. We ran and analyzed a set of ten hydrodynamical simulations of the formation of embedded star clusters through the collapse of turbulent massive molecular clouds. We systematically studied the fractality of our star clusters, showing that they are all extremely substructured (fractal dimension $D=1.0-1.8$). We also found that fractality is slowly reduced, with time, on small scales, while it persists on large scales on longer timescales.
Signatures of rotation are found in different simulations at every time of the evolution, even for slightly supervirial substructures, proving that the parent molecular gas transfers part of its angular momentum to the new stellar systems.
\end{abstract}




\section{Introduction}\label{intro}

The theoretical study of star clusters needs a more realistic description of their initial conditions. There are stronger and stronger indications that the properties of these systems could be the imprints of their formation process. For example, young massive clusters and open clusters show substrutctures with complex kinematics and fractality \citep[e.g.,][]{Cartwright04, Sanchez09, Parker12, Kuhn19, Cantat-Gaudin19}, traces of ongoing dispersal \citep[e.g.,][]{Kuhn19,Cantat-Gaudin19}, believed to be due to sudden gas expulsion \citep[e.g.,]{Tutukov78,Lada84,Geyer01, Baumgardt07} and indications of rotation \citep[e.g.,][]{Henault-Brunet12}. 

Understanding the formation of star clusters in the local Universe might also be crucial to explain the properties of older systems as globular clusters. Signatures of rotation are also found for these more evolved systems \citep[e.g.,][]{vanLeeuwen00,Pancino07,Bianchini13, Fabricius14, Kamann18}. Furthermore, globular clusters show multiple populations of stars with slightly different chemical properties, kinematics and segregation \citep[see][and references therein]{Gratton04,Marino08, Milone10, Carretta11, Gratton12,Bastian18, Milone20}.  These distinct populations might be explained by a non-monolithic formation of their host cluster \citep{Perets14,Mastrobuono-Battisti16,Gavagnin16,Bekki16,Mastrobuono-Battisti19}. 

The degree of fractality (i.e., of substructuring) and the amount of rotation of a young star cluster are both expected to have an impact on its dynamical evolution, particularly at the early stages of its assembly. In particular, they are both enhancing the local probability of two-body encounters, e.g., shortening the two-body relaxation of such systems. 
This impact on the relaxation timescale is particularly important because dynamics is predicted to play a crucial role, for example, in the formation of intermediate-mass black holes \citep{Colgate67, Ebisuzaki01,PortegiesZwart04,Freitag06,Giersz15,Mapelli16} and the formation and evolution of massive binary systems, which might be the precursors of compact object mergers \citep{Ziosi14,Kimpson16,Mapelli16,Banerjee17,Fujii17,Banerjee18a,Banerjee18b,DiCarlo19,Kumamoto19}.  

Fractality has been quite widely investigated for observed systems and through pure \textit{N}-body simulations. However, few studies focused on the fractality of star clusters forming in hydrodynamical simulations of the collapse of turbulent molecular clouds. A first attempt was performed by \citet{Schmeja06}. These authors tested the so-called $Q$ parameter, first defined by \citet{Cartwright04}, on both observations of young embedded star clusters and the smoothed-particle hydrodynamics (SPH) simulations by \citet{Schmeja04}. They reported values of $Q$ comparable between their models and real star clusters and  found no significant correlation between the fractality of their sink particle distributions and the properties of the turbulent field induced in the simulated collapsing molecular clouds. A similar attempt can be found in \citet{Maschberger10}, who analyzed two simulations with cloud mass equal to $10^3$ and $10^4$~M$_{\odot}$, performed by \citet{Bonnell03} and \citet{Bonnell08}, respectively. The resulting sink particle clusters were both forming with low values of $Q$ (of the order of $0.4-0.5$), typical of a very high degree of substructuring. However, for the lower mass cluster, which is bound, $Q$ evolves to values typical of no fractality in around a couple of free-fall times, while for the higher mass cluster, which is initially unbound, $Q$ stays more or less constant. This is a proof that star clusters assembling hierarchically are expected to form fractal and then to ``lose'' substructures due to their mergers and to relaxation processes.

Low values of $Q$ at the early stage of star formation have also been found by \citet{Girichidis12}, though for much smaller (100~$M_{\odot}$) and strongly unstable clouds. \cite{Girichidis12} also found a possible dependence (though quite mild) of $Q$ on both the initial density profile and mode of turbulence of the collapsing cloud. Finally, no strong dependence of $Q$  on  stellar feedback was found in the hydrodynamical simulations analyzed in \citet{Parker15} and \citet{Gavagnin17}. In general, an evolution of $Q$ from small values to larger values (no fractality) is found also in these studies.

Rotation in embedded star (proto-)clusters forming in hydrodynamical simulations has been way less investigated in literature. Recent studies by \citet{Lee16} and \citet{Mapelli17} showed that such star clusters inherit significant rotation from their parent cloud, by large-scale torques from the gas and from angular momentum conservation in the collapse of the densest cores.
Indeed, possible signatures of rotation are found in observations (through different gas tracers) of collapsing molecular clouds, sometimes at large scales \citep[e.g.,][]{Galvan-Madrid09,Li17}, but most importantly at the loci of convergence of different turbulence-induced filaments \citep{Ho86, Zhang97, Liu12, Beuther13, Dalgleish18, Juarez19, Liu19, Trevino-Morales19}.

In this paper, we study these two properties for ten hydrodynamic simulations of the formation of star clusters by the collapse of massive molecular clouds. A brief description of the initial conditions and methods adopted to run our set can be found in Section \ref{meth}. In Section \ref{fracres}, we present our analysis of fractality, performed through the adoption of few different diagnostic indicators. The analysis of rotation can be found in Section \ref{rotres}. A discussion of our results is presented in Section \ref{discus}. We summarize our results in Section \ref{summary}.

\begin{figure*}
\begin{center}
\includegraphics[scale=0.75,trim={0 2cm 0 2cm},clip]{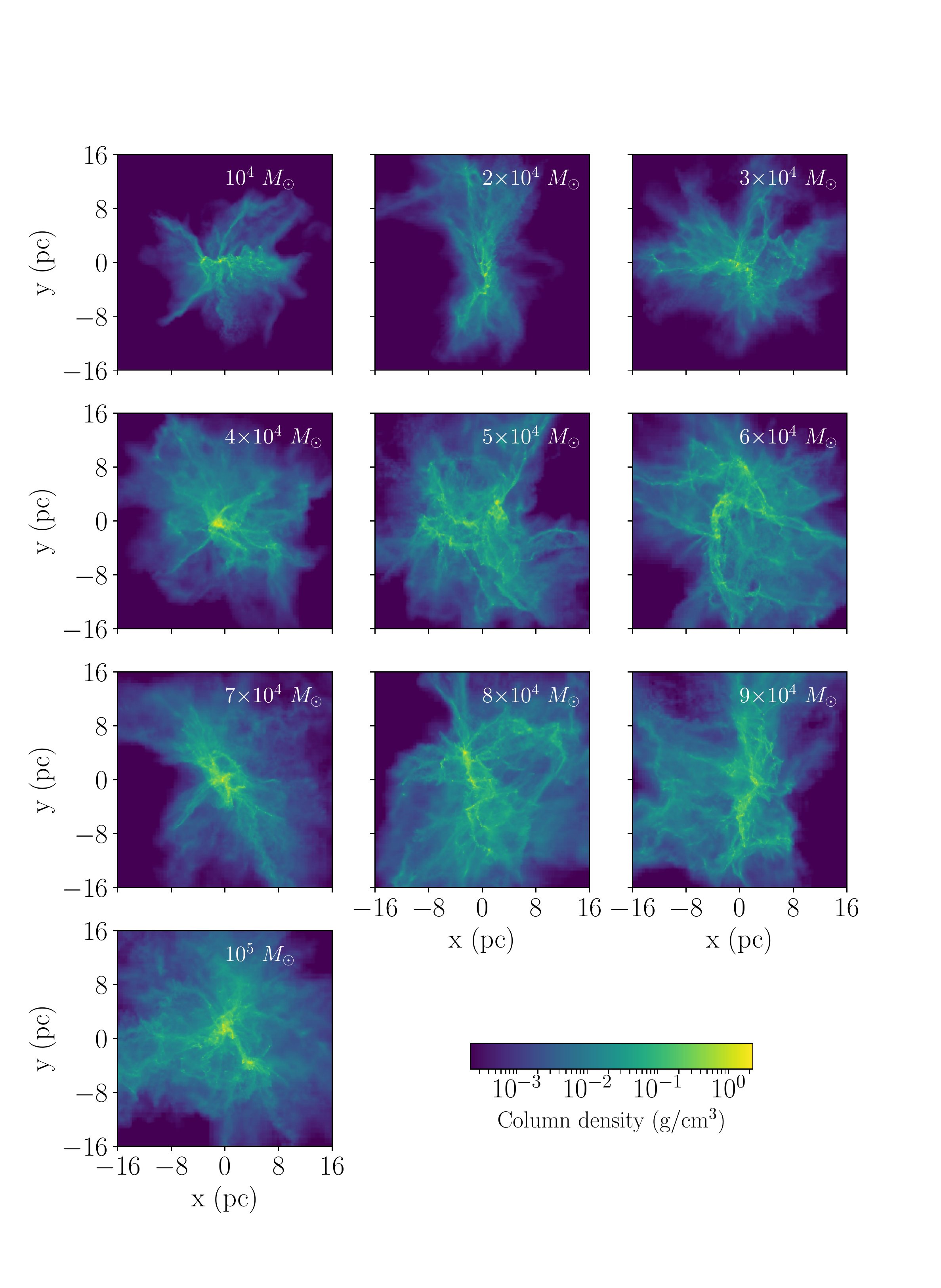}
\caption{Projection map of the gas particle density at $t_{sim}=3$~Myr, for all the simulations of our set. 
}\label{map_gas}
\end{center}
\end{figure*}

\begin{figure*}
\begin{center}
\includegraphics[scale=0.75,trim={0 2cm 0 2cm},clip]{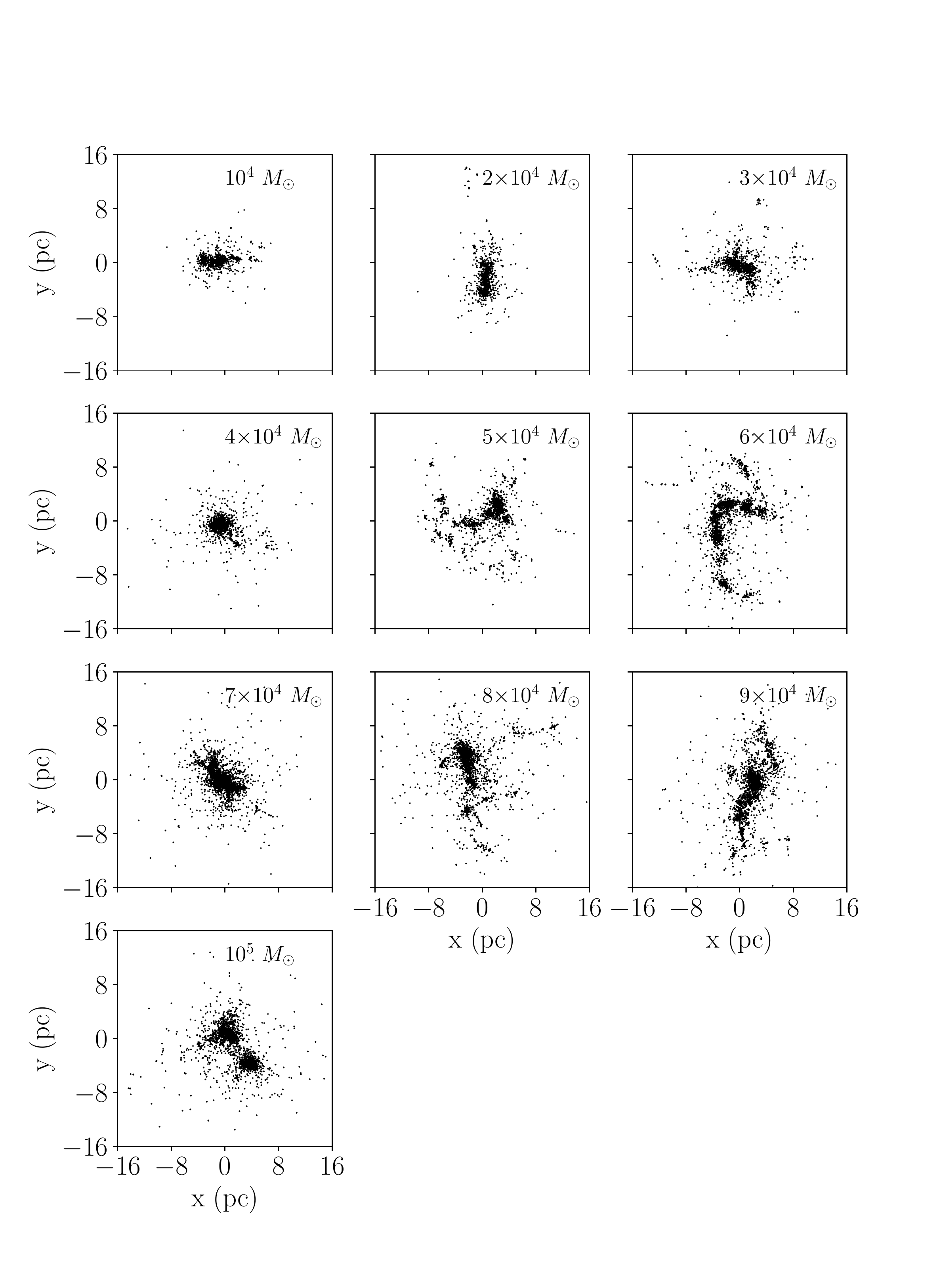}
\caption{Projection of the sink particle distribution at $t_{sim}=3$~Myr, for all the simulations of our set.
}\label{map_sinks}
\end{center}
\end{figure*}

\begin{figure*}
\begin{center}
\includegraphics[scale=0.5]{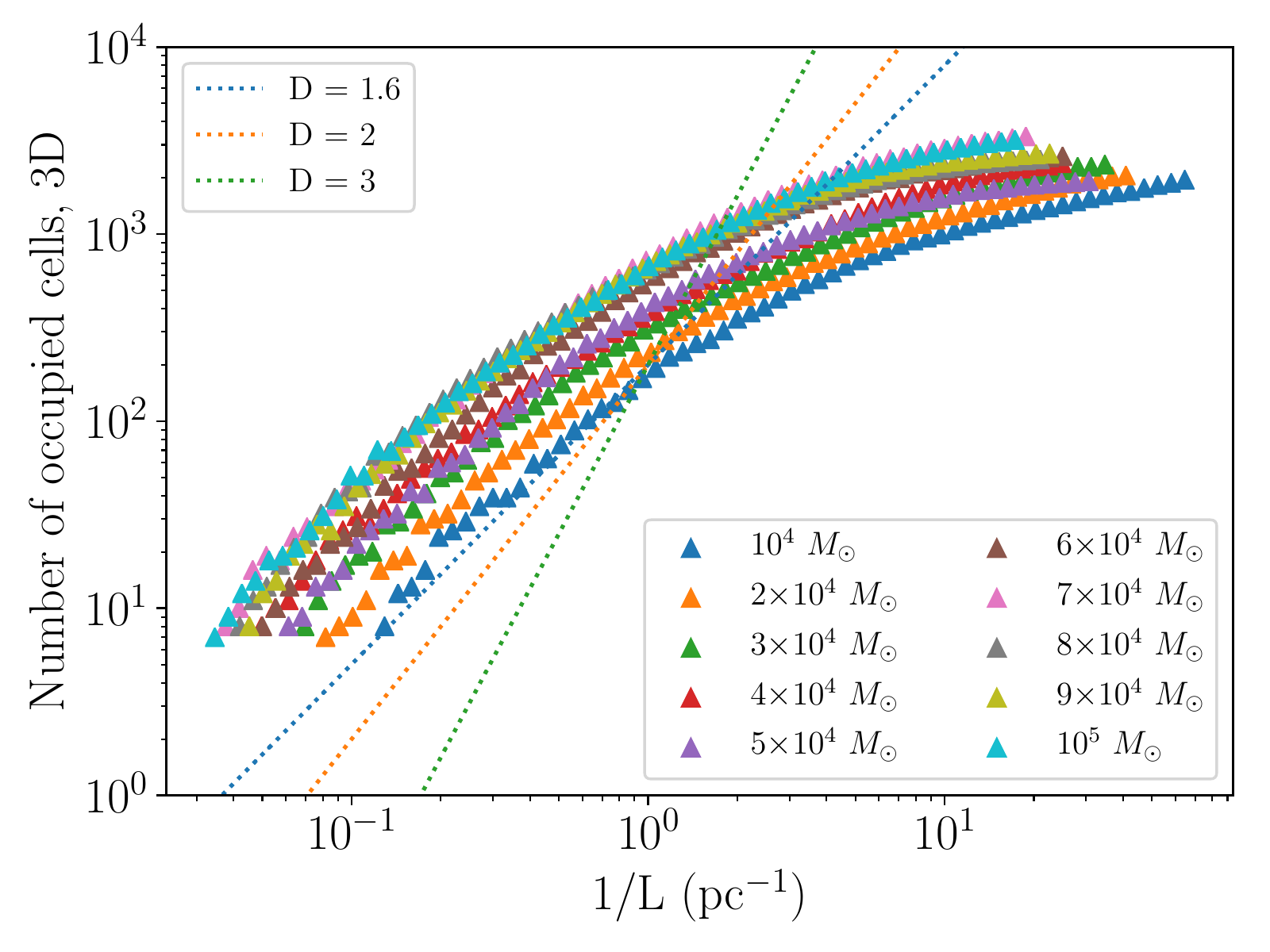}
\includegraphics[scale=0.5]{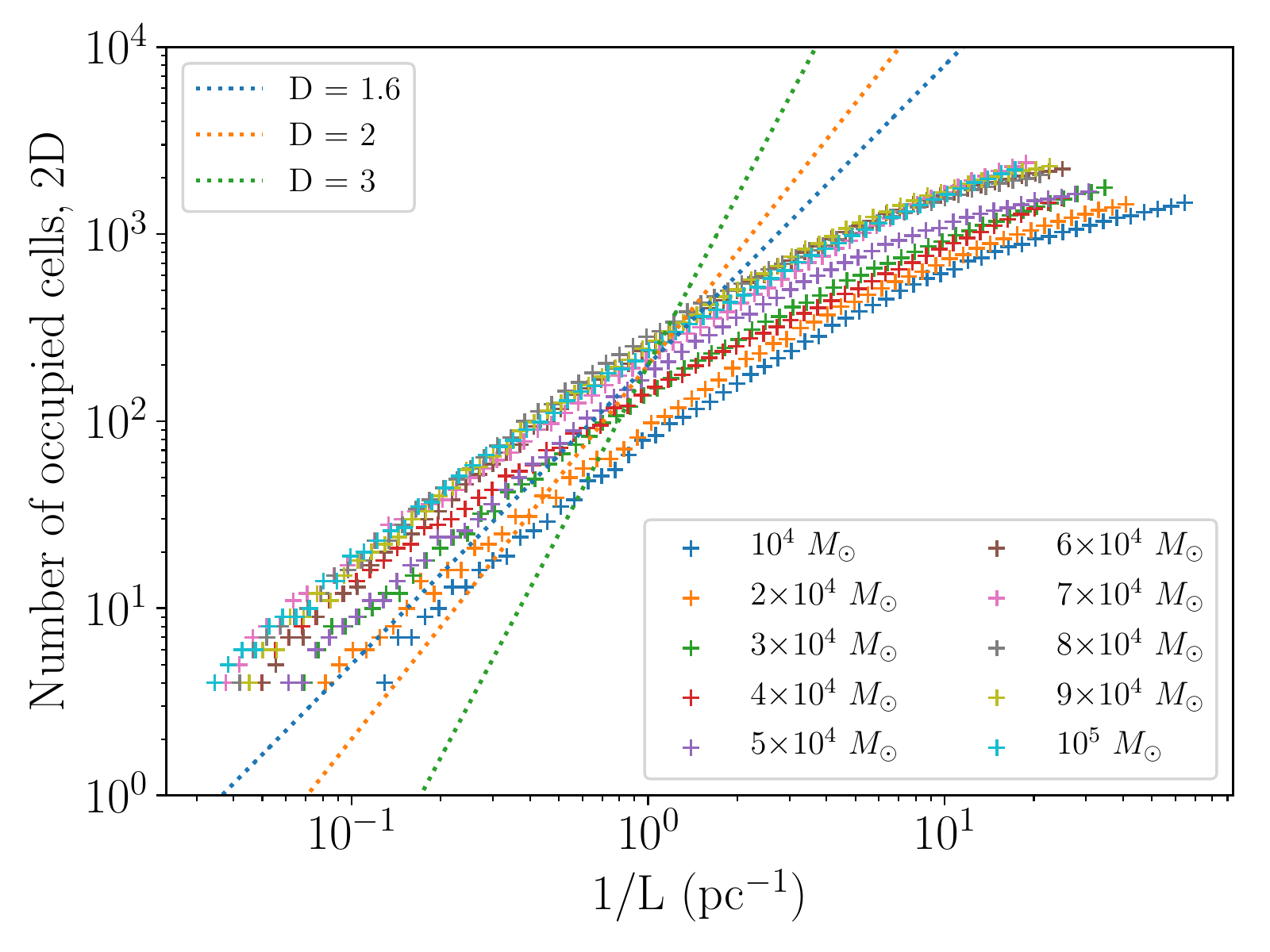}
\caption{Fractal dimension, as calculated through the `box-counting' method, for the 3D distribution (left) and a 2D projection (right) of the sink position. The plot shows the number of cells (with size $L$) occupied by at least one sink, for 3D/2D grids covering the sink clusters (see text for a detailed explanation). The slope of such logarithmic curves is equal to the fractal dimension $D_B$. The three dotted lines show the ideal curves for $D_B=1.6,2.0,3.0$.
}\label{frac_dim_B}
\end{center}
\end{figure*}

\begin{figure*}
\begin{center}
\includegraphics[scale=0.5]{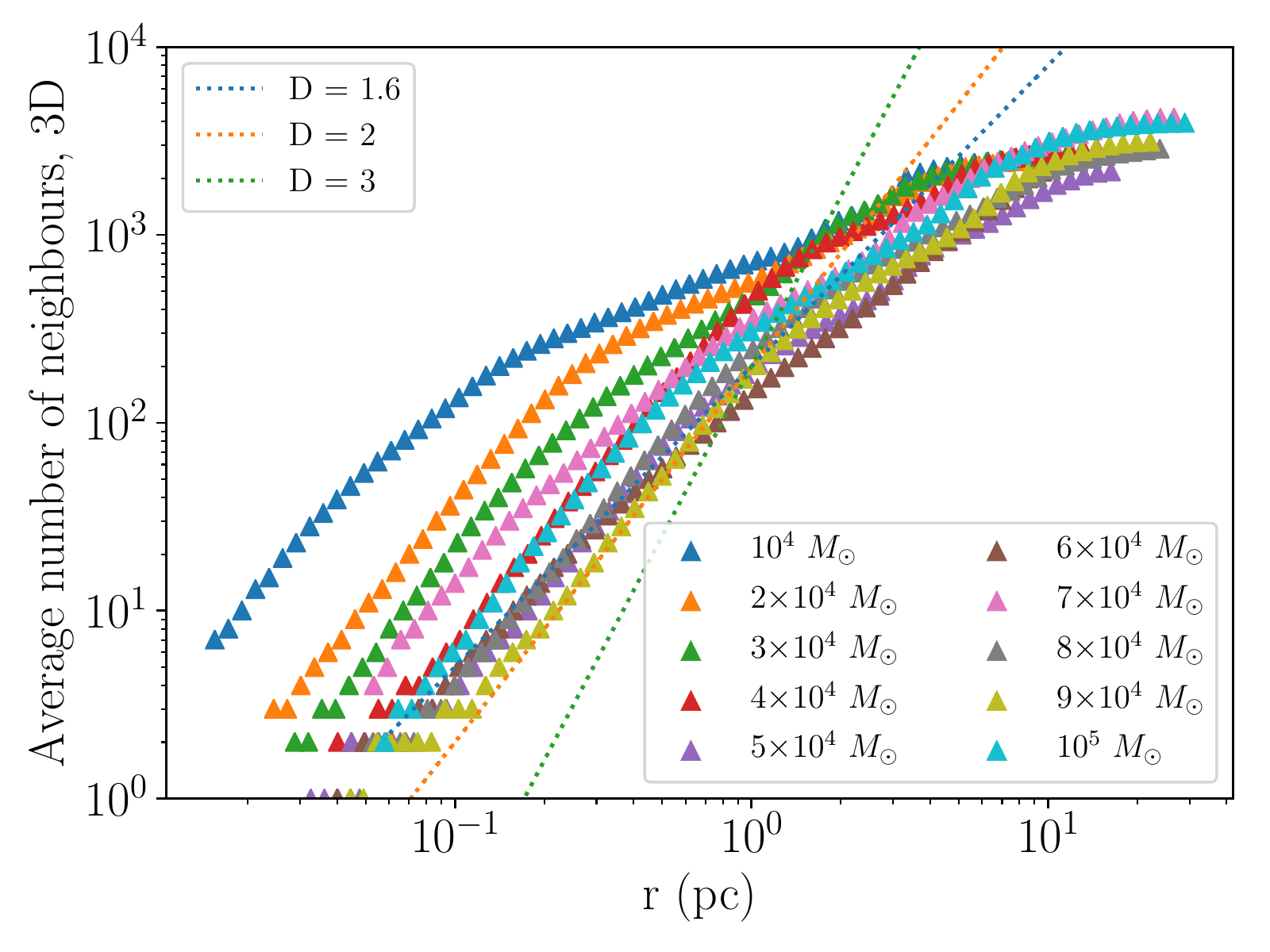}
\includegraphics[scale=0.5]{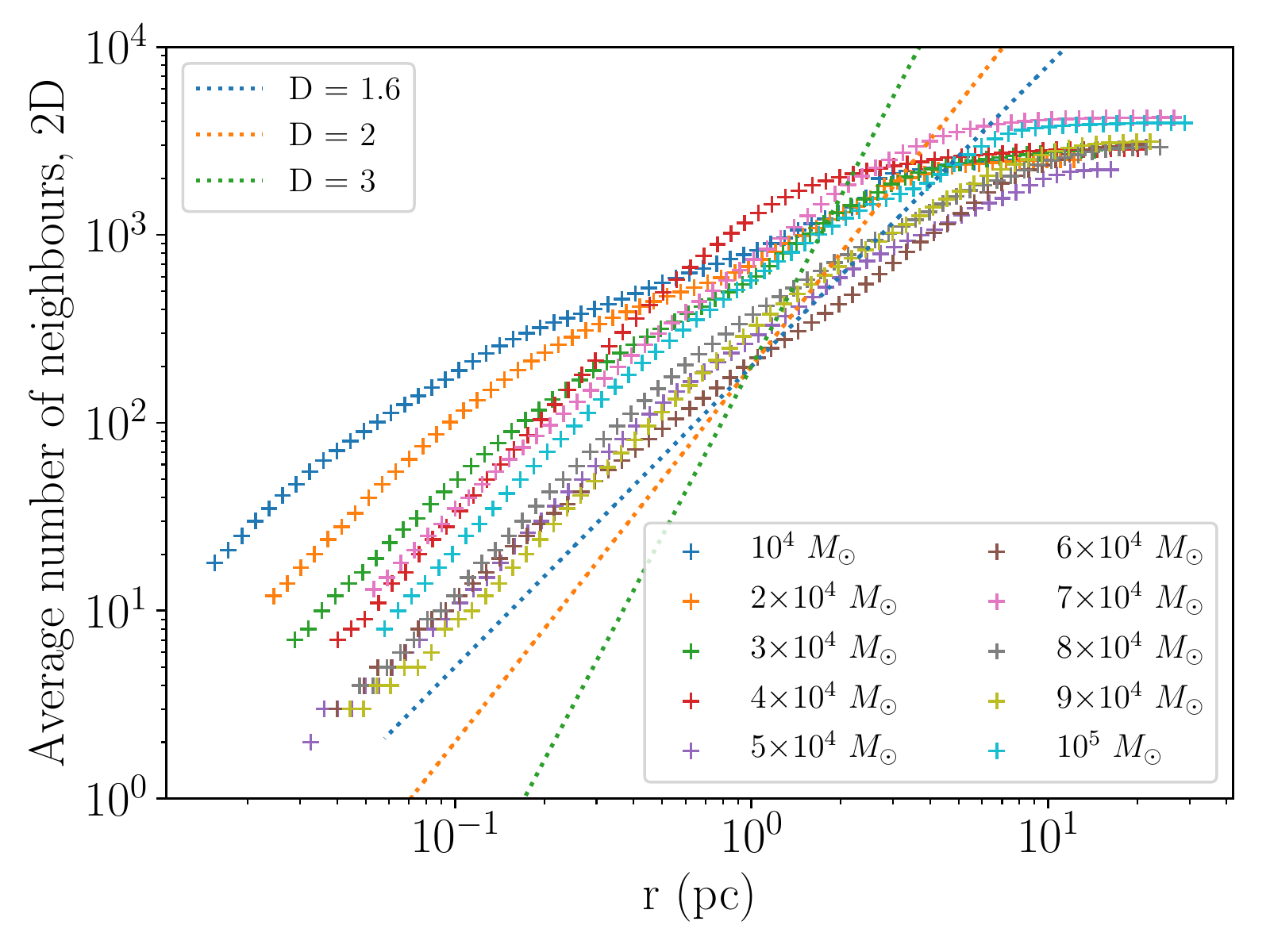}
\caption{Fractal dimension, as calculated through the `neighbour-counting' method, for the 3D distribution (left) and a 2D projection (right) of the sink position. The plot shows the number of sinks contained in a circle/sphere of radius $r$ (see text for a detailed explanation). The slope of such logarithmic curves is equal to the fractal dimension $D_C$. The three dotted lines show the ideal curves for $D_C=1.6,2.0,3.0$. 
}\label{frac_dim_C}
\end{center}
\end{figure*}

\section{Hydrodynamical simulations}\label{meth}

We analyzed 10 hydrodynamical simulations of turbulent molecular clouds performed with the SPH code GASOLINE \citep{Wadsley04,Wadsley17}. We adopted an adiabatic equation of state, coupled to the radiative cooling algorithm described in \citet{Boley09} and \citet{Boley10}. The cooling is calculated from the divergence of the flux $\nabla \cdot F=-(36\pi)^{1/3}s^{-1}\sigma(T^4-T_{irr}^4)(\Delta\tau+1/\Delta\tau)^{-1}$, where $\sigma$ is the Stefan-Boltzmann constant, $T_{irr}$ represents the incident radiation, $s=(m/\rho)^{1/3}$ and $\Delta\tau=sk\rho$, with $m$ and $\rho$ being the particle mass and density and $k$ being the local opacity. For $k$, tabulated values of the Planck and Rosseland dust opacities are used, taken from \citet{Dalessio01}, while we adopted $T_{irr}=10$ K.

The clouds have an initial uniform density and temperature of 250~$\mathrm{cm^{-3}}$ and 10~K, respectively, distributed on a sphere with mass equal to $10^4-10^5$~M$_{\odot}$ (see Table \ref{properties}). This leads to a fixed cloud free-fall time of $t_{ff}=2$~Myr.

All the clouds are initially turbulent, so to be in a marginally bound state. This means that their virial ratio $\alpha_{vir}=T/|V|=1$, where $T$ and $V$ are the kinetic and potential energy, respectively. The turbulence consists of a divergence-free Gaussian random velocity field, following a \citet{Burgers48} power spectrum. The turbulence seed is different for each simulation.

All the simulations have a fixed number of $10^7$ gas particles, corresponding to a mass resolution of $10^{-3}-10^{-2}$~M$_\odot$, with gravitational softening of $\epsilon=10^{-4}$~pc.

We modelled the formation of stars through a sink particle algorithm adopting the same criteria as in \citet{Bate95}. In particular, we adopted a density threshold\footnote{The main difference between our simulation setup and the one of \cite{Mapelli17} is the density threshold, being $10^5$ cm$^{-3}$ in \cite{Mapelli17}. This difference affects the initial virial state of the simulated sink clusters: our star clusters end up having $\alpha_{vir}\gtrsim{}0.5$, while the clusters simulated by \cite{Mapelli17} are sub-virial. This deserves further analysis and should be kept into account when interpreting the results of numerical studies based on sink particle algorithms.} of $10^{7}$ cm$^{-3}$ and we set a sink radius of $r_{s}=2\times 10^{-3}$ pc.

\begin{table*}
\caption{Properties of the simulated star clusters at $t_{sim}=3$~Myr.}
\label{properties}
\centering  
\begin{tabular}{l l l l l l l l l l l}
\hline
 $M_{mc}$ (M$_{\odot}$) & $R_{mc}$ (pc) & $N_{s}$ & $M_{s}$ (M$_{\odot}$) & $\epsilon_{sf}$ & $D_{B3D, s}$ & $D_{B2D, s}$ & $D_{C3D, s}$ & $D_{C2D,s}$ & $Q_{2D,s}$ & $Q_{3D,s}$ \\

\hline
 
$10^4$ & 5.4 & 2531 & $4.22\times 10^3$ & 0.42 & 1.38$\pm$0.03 & 1.20$\pm$0.02 & 1.04$\pm$0.04 & 0.87$\pm$0.03 & 0.40 & 0.24 \\ 
$2\times 10^4$ & 6.8 & 2571 & $6.69\times 10^3$ & 0.33 & 1.40$\pm$0.03 & 1.19$\pm$0.02 & 1.38$\pm$0.04 & 1.06$\pm$0.03 & 0.40 & 0.25 \\ 
$3\times 10^4$ & 7.8 & 2825 & $1.03\times 10^4$ & 0.34 & 1.34$\pm$0.03 & 1.21$\pm$0.02 & 1.51$\pm$0.03 & 1.22$\pm$0.02 & 0.49 & 0.31 \\ 
$4\times 10^4$ & 8.6 & 2868 &$1.44\times 10^4$ & 0.36 & 1.30$\pm$0.03 & 1.04$\pm$0.02 & 1.76$\pm$0.02 & 1.69$\pm$0.01 & 0.76 & 0.32 \\ 
$5\times 10^4$ & 9.2 & 2231 & $1.41\times 10^4$ & 0.28 & 1.48$\pm$0.03 & 1.34$\pm$0.02 & 1.51$\pm$0.02 & 1.40$\pm$0.02 & 0.35 & 0.20 \\ 
$6\times 10^4$ & 9.8 & 3054 & $2.04\times 10^4$ & 0.34 & 1.52$\pm$0.02 & 1.33$\pm$0.02 & 1.43$\pm$0.02 & 1.25$\pm$0.01 & 0.37 & 0.21 \\ 
$7\times 10^4$ & 10.3 & 4214 & $3.15\times 10^4$ & 0.45 & 1.40$\pm$0.02 & 1.12$\pm$0.01 & 1.30$\pm$0.03 & 1.34$\pm$0.01 & 0.69 & 0.26 \\ 
$8\times 10^4$ & 10.8 & 2945 & $2.83\times 10^4$ & 0.35 & 1.56$\pm$0.04 & 1.30$\pm$0.01 & 1.61$\pm$0.03 & 1.42$\pm$0.02 & 0.46 & 0.28 \\ 
$9\times 10^4$ & 11.2 & 3161 & $3.05\times 10^4$ & 0.34 & 1.49$\pm$0.04 & 1.27$\pm$0.02 & 1.60$\pm$0.02 & 1.48$\pm$0.02 & 0.43 & 0.27 \\ 
$10^5$ & 11.6 & 3944 & $3.80\times 10^4$ & 0.38 & 1.40$\pm$0.03 & 1.17$\pm$0.01 & 1.49$\pm$0.04 & 1.41$\pm$0.02 & 0.53 & 0.28 \\ 
\hline
\end{tabular}

\footnotesize{Column~1 ($M_{mc}$): initial molecular cloud mass; column~2 ($R_{mc}$): initial molecular cloud radius; column~3 ($N_s$): number of sink particles formed after $1.5~t_{\rm ff}$; column~4 ($M_s$): total mass of sink particles after $1.5~t_{\rm ff}$; column~5 ($\epsilon_{sf}$): star formation efficiency at $1.5~t_{\rm ff}$; column~6 ($D_{B3D,s}$): 3D fractal dimension of sink particles calculated with the box-counting method (see the text for details); column~7 ($D_{B2D,s}$): 2D fractal dimension of sink particles calculated with the box-counting method; column~8 ($D_{C3D,s}$): 3D fractal dimension of sink particles calculated with the neighbour-counting method (see the text for details); column~9 ($D_{C2D,s}$): 2D fractal dimension of sink particles calculated with the neighbour-counting method; column~10 ($Q_{2D,s}$): $Q$ parameter of sink particles calculated in two dimensions;  column~11 ($Q_{3D,s}$): $Q$ parameter of sink particles calculated in three dimensions.}
\end{table*}

\begin{figure}
\begin{center}
\includegraphics[scale=0.5]{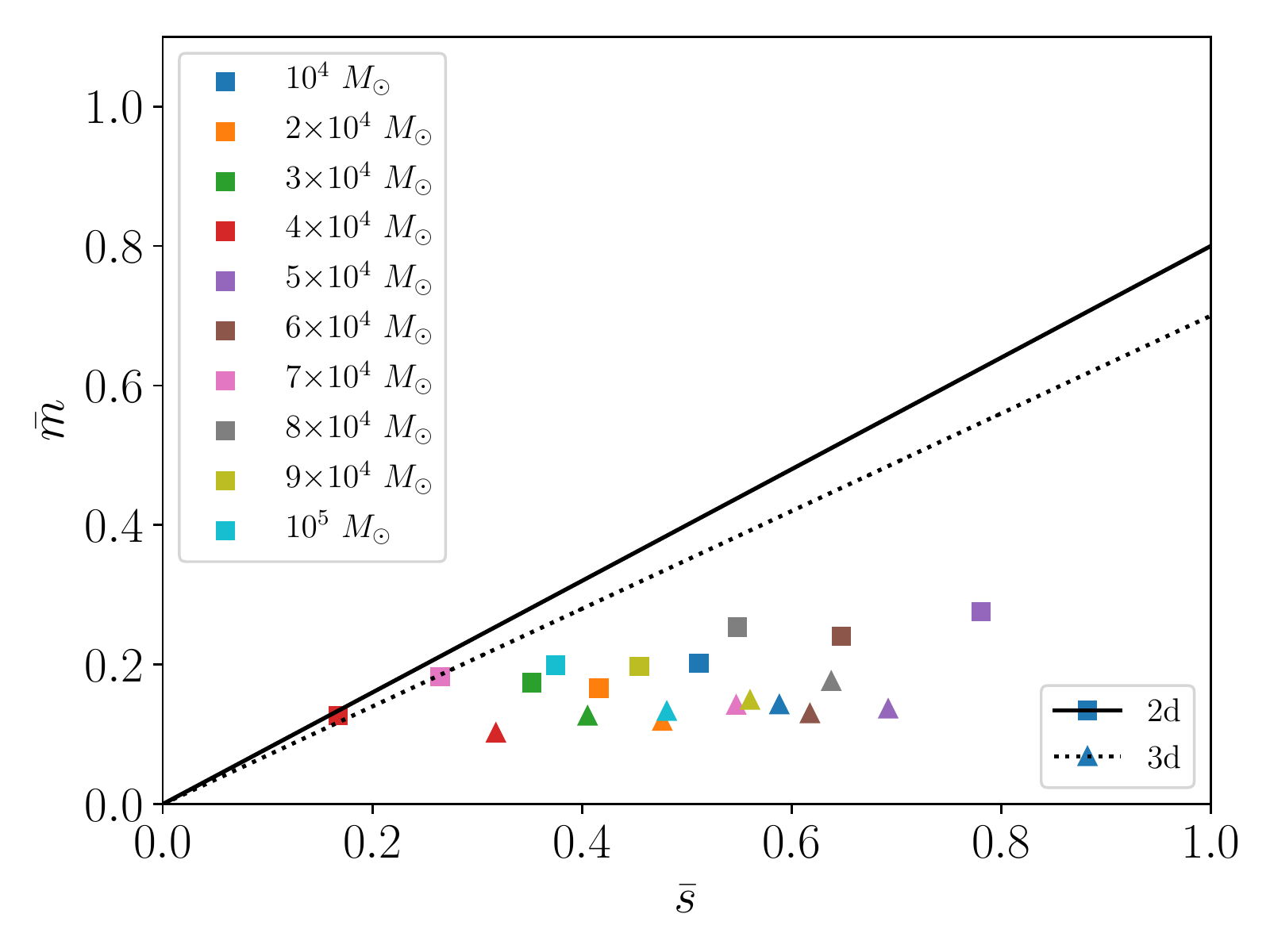}
\caption{Mean inter-particle distance $\bar s$ versus mean edge length of the minimum-spanning-tree $\bar m$ (see text for details) for all the simulations in our set at $t_{sim}=3$~Myr. The squares and solid line are relative to the 2D calculation, while the triangles and dotted line are relative to the 3D calculation.
}\label{Q}
\end{center}
\end{figure}

\begin{figure}
\begin{center}
\includegraphics[scale=0.5]{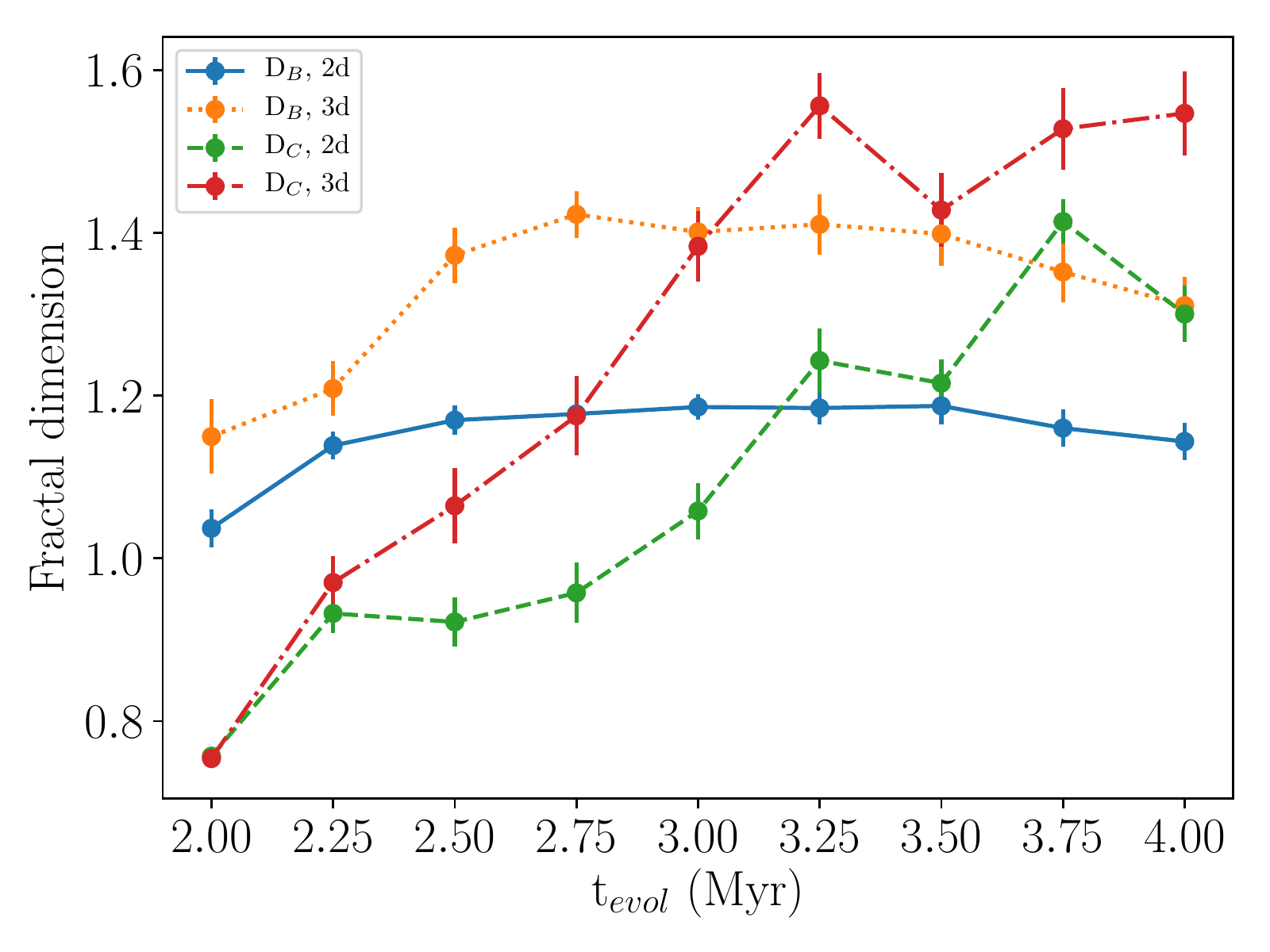}
\caption{Evolution of the fractal dimension for $M_{cl}=2\times 10^4$~M$_\odot$. The results for the `box-counting' method are shown by the solid  (2D) and dotted (3D) lines, while those for the `neighbour-counting' method are shown by the dashed (2D) and dashed-dotted (3D) lines. 
}\label{D_evo}
\end{center}
\end{figure}

\begin{figure}
\begin{center}
\includegraphics[scale=0.53]{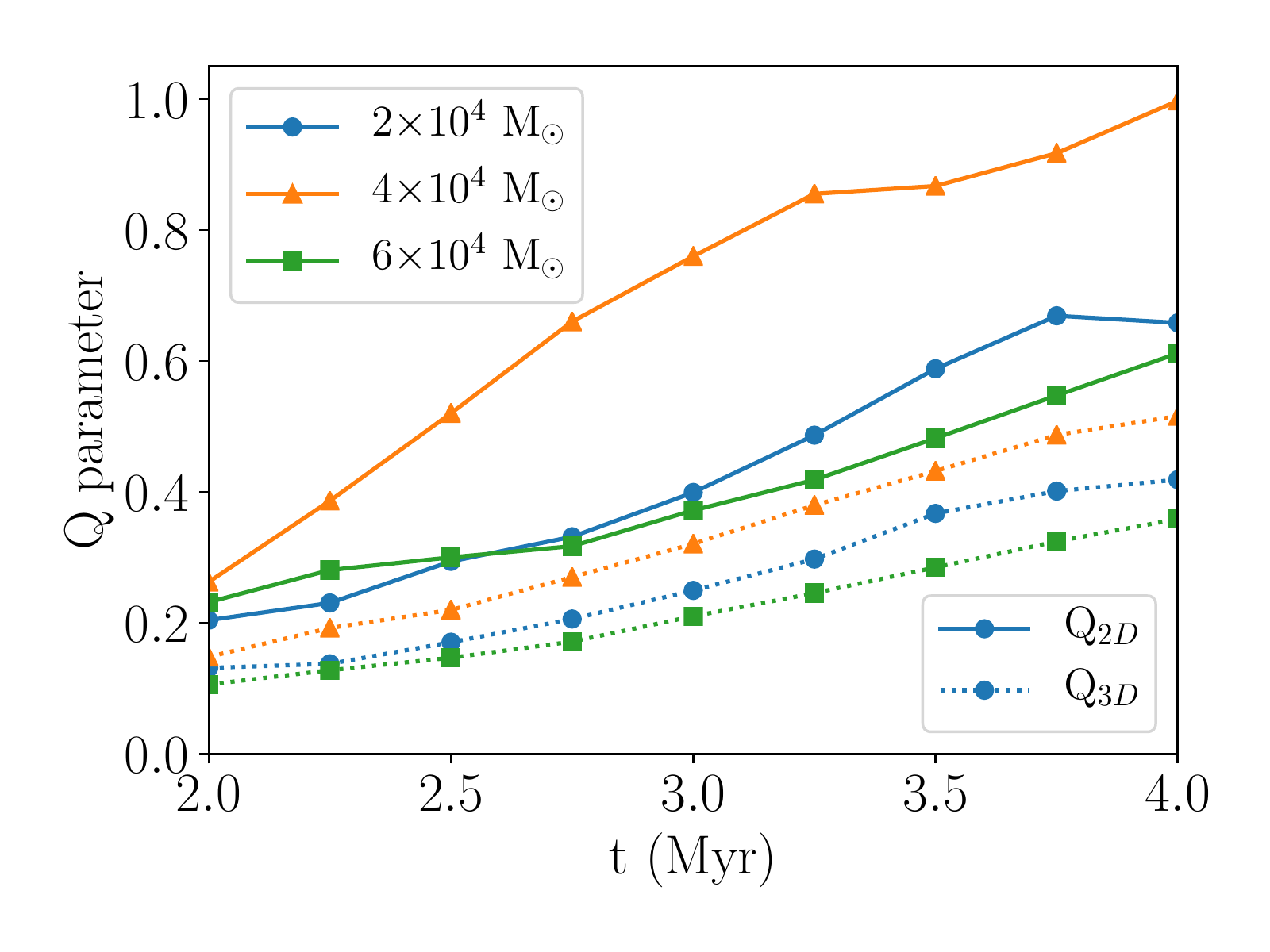}
\caption{Evolution of the $Q$ parameter for $M_{cl}=2, 4, 6 \times 10^4$~M$_{\odot}$ (circles, triangles and squares, respectively). The solid and dotted lines show the results for the 2D and 3D calculation, respectively.}\label{Q_evo}
\end{center}
\end{figure}

\section{Results}\label{results}

Unless differently stated, we analyze the simulations at an evolutionary time of $t_{sim}=3$ Myr (i.e., equal to 1.5 times the original cloud free-fall time $t_{cl,ff}$), at which we investigated the main properties of the sinks, as summarized in Table \ref{properties}. This choice is somewhat arbitrary, but it is roughly consistent with the time at which we expect stellar feedback to start expelling gas from the parent cloud and lead to a saturation of star formation. As shown in Table \ref{properties}, the sink formation efficiency, defined as $\epsilon_{sf}=M_{s}/M_{mc}$ (where $M_{mc}$ is the initial cloud mass and $M_s$ is the mass in sinks), ranges between 0.28 and 0.45, consistent with previous simulations, that also considered stellar feedback \citep[e.g.,][]{VazquezSemadeni10,Dale14, Gavagnin17, Li19}.

Figure \ref{map_gas} shows projection maps of the gas density at the end of the simulations, while Figure \ref{map_sinks} shows the distribution of the sinks formed in each cloud.

In the following, we will focus on some major structural and kinematic properties of the formed star clusters and of their sub-clumps.

\begin{figure*}
\begin{center}
\includegraphics[scale=0.33]{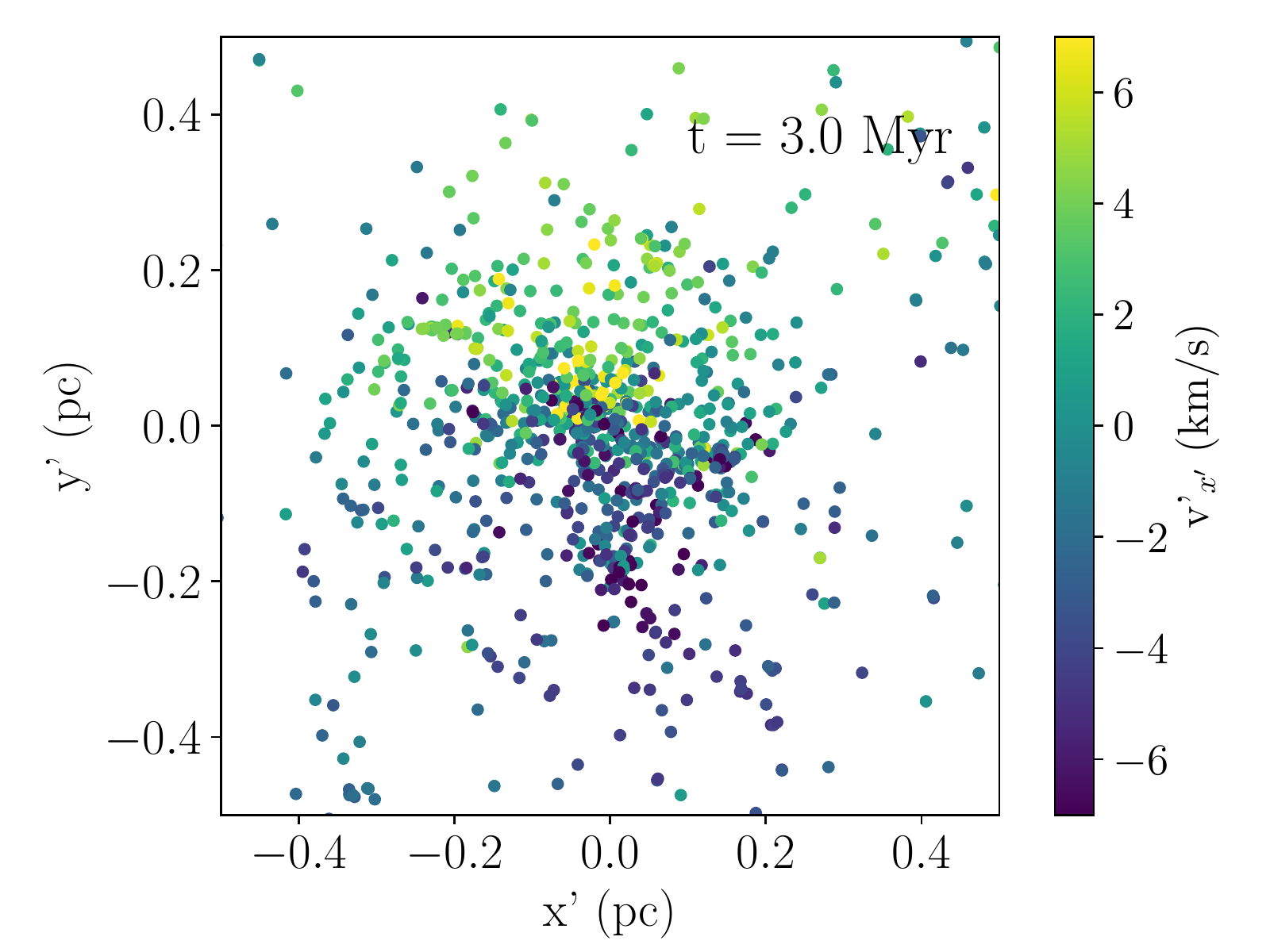}
\includegraphics[scale=0.33]{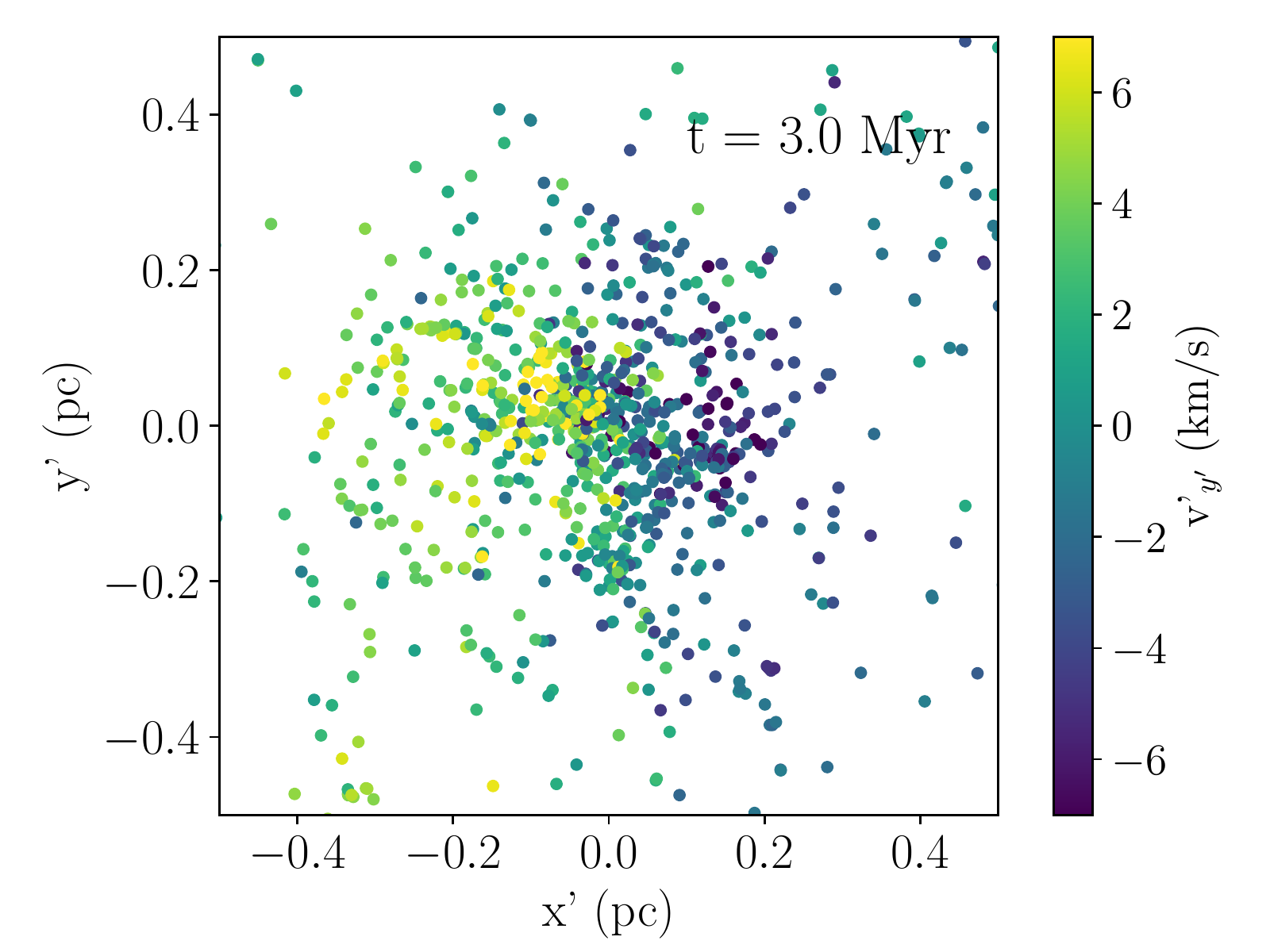}
\includegraphics[scale=0.33]{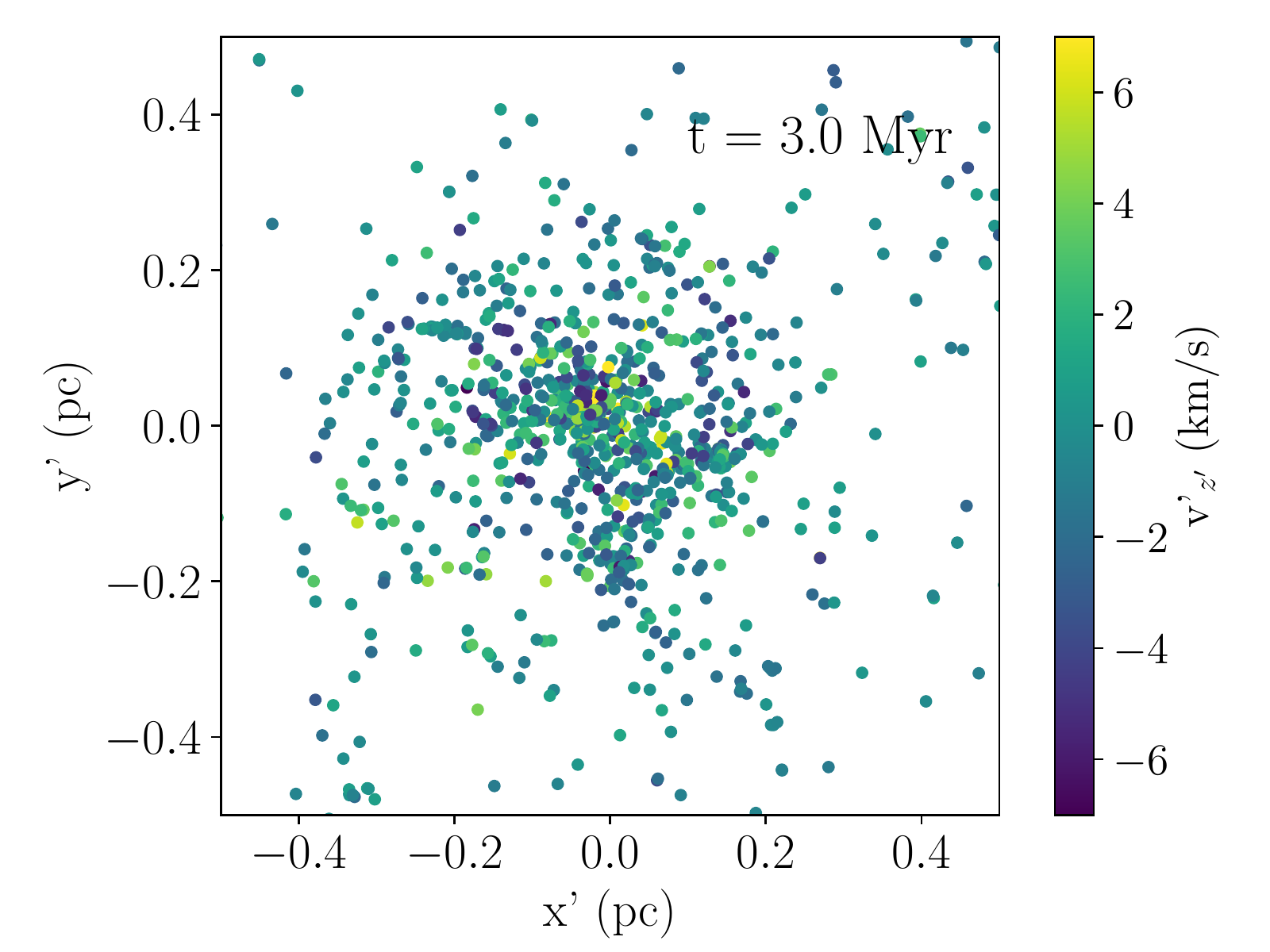}
\includegraphics[scale=0.33]{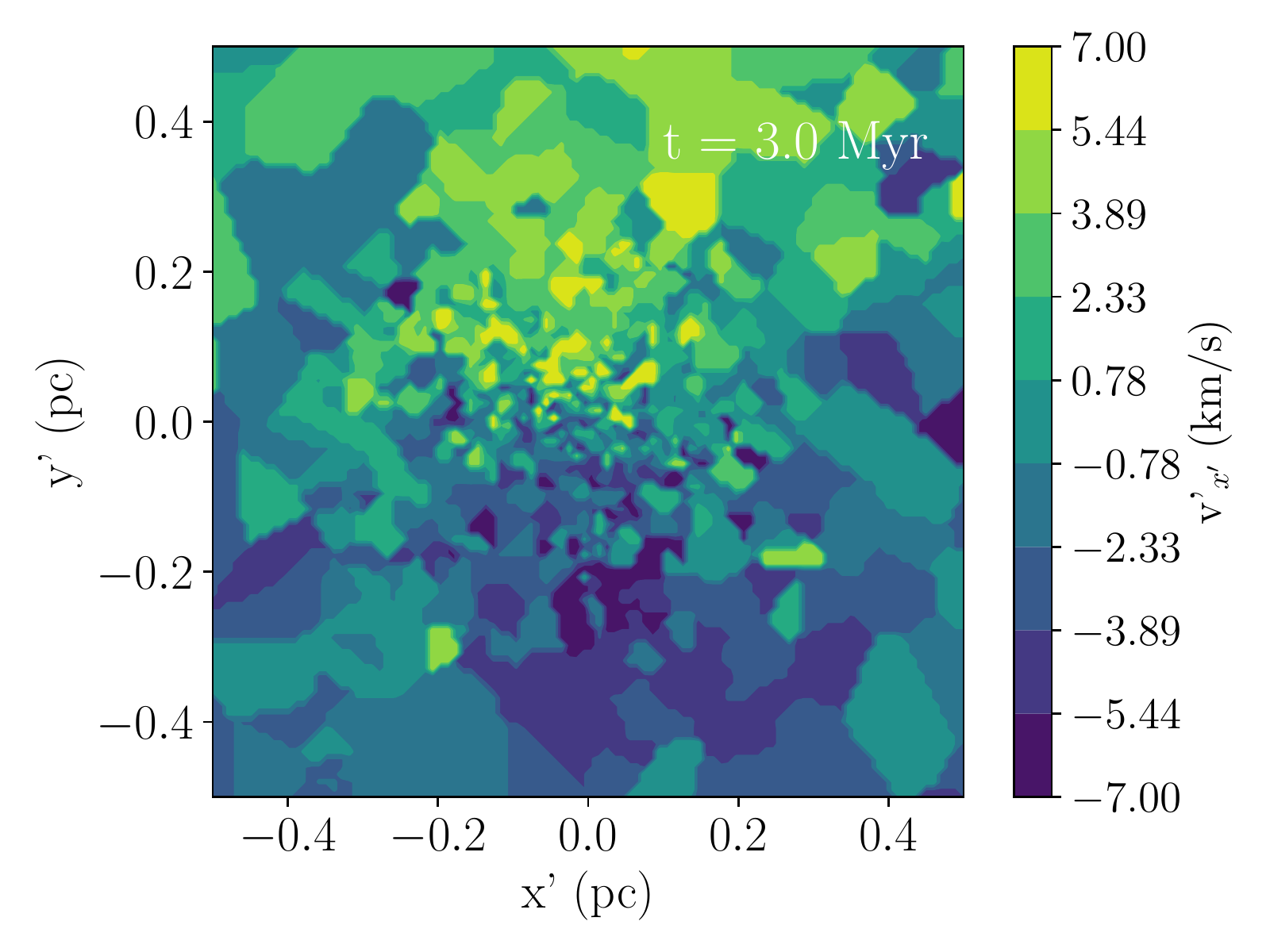}
\includegraphics[scale=0.33]{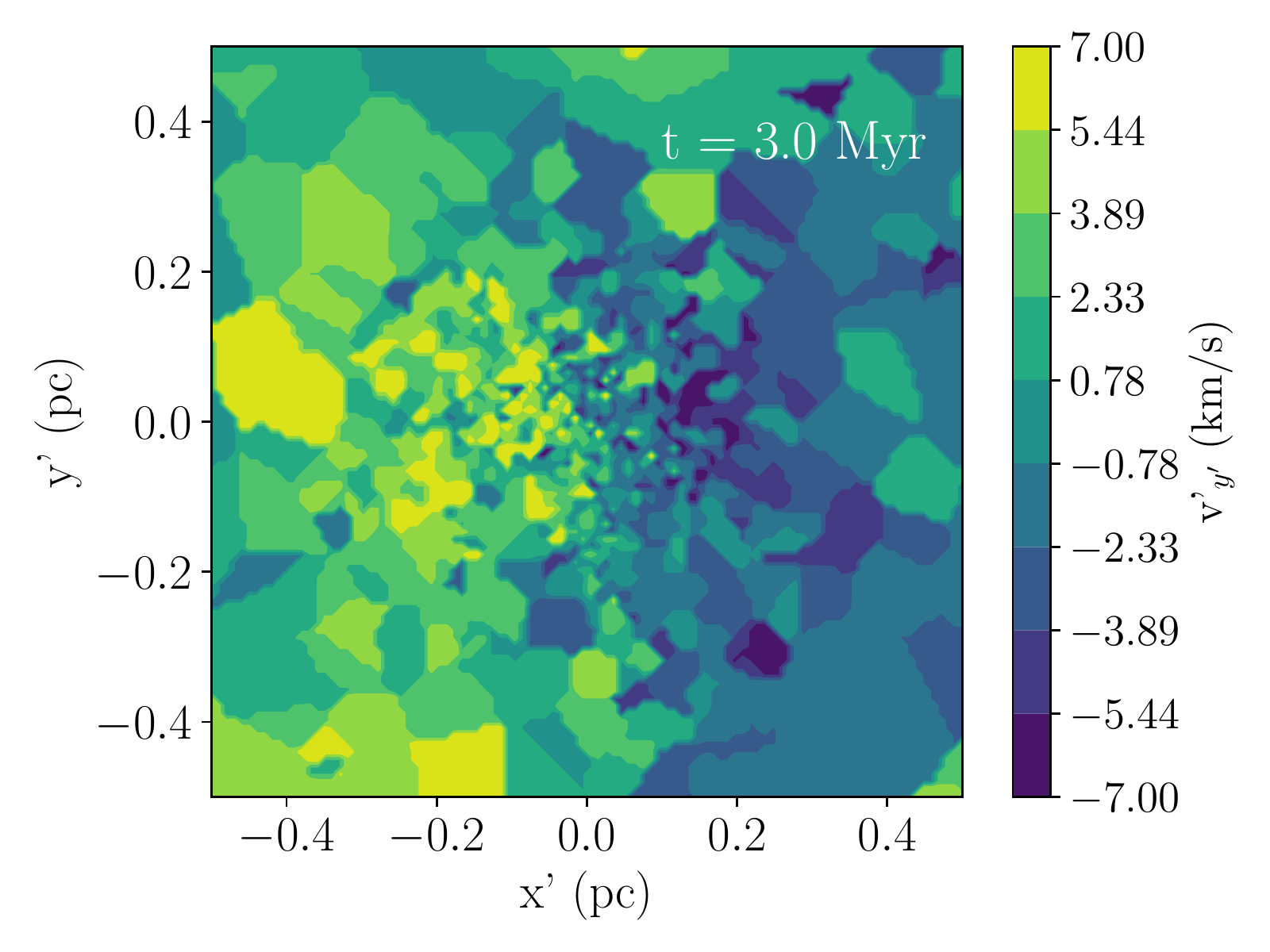}
\includegraphics[scale=0.33]{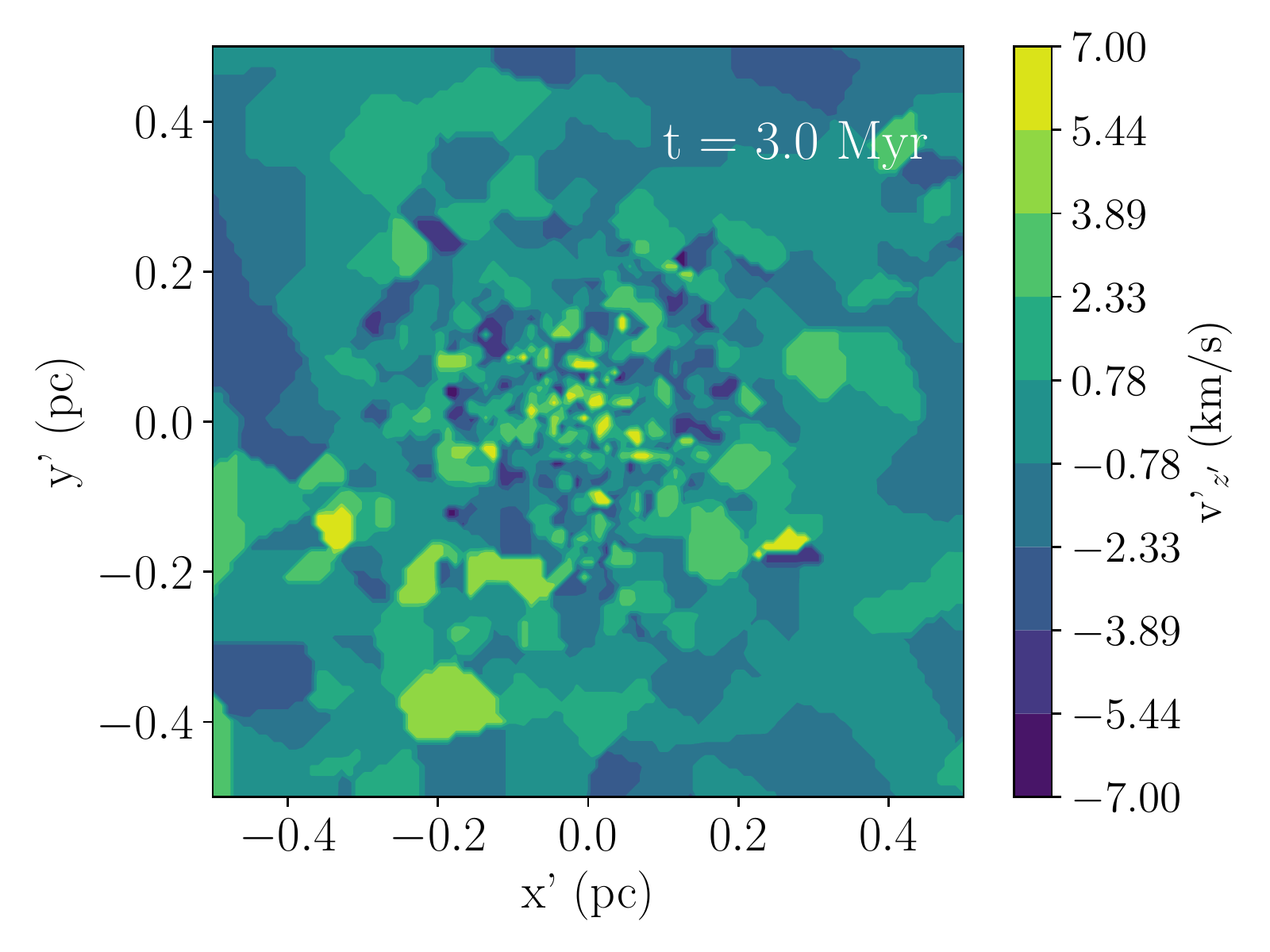}
\caption{Scatter plots (upper panels) and Voronoi tessellation maps (lower panels) of the highest angular momentum sub-cluster formed in the $M_{cl}=2\times 10^4$~M$_{\odot}$ simulation, at 3~Myr. We show the $x'-y'$ plane, where $z'$ is the direction of the angular momentum of the sub-cluster. The colour map refers to the three components of the velocity: $v'_{x'}$ (left), $v'_{y'}$ (center) and $v'_{z'}$ (right).  
}\label{rot_sink}
\end{center}
\end{figure*}

\begin{figure*}
\begin{center}
\includegraphics[scale=0.33]{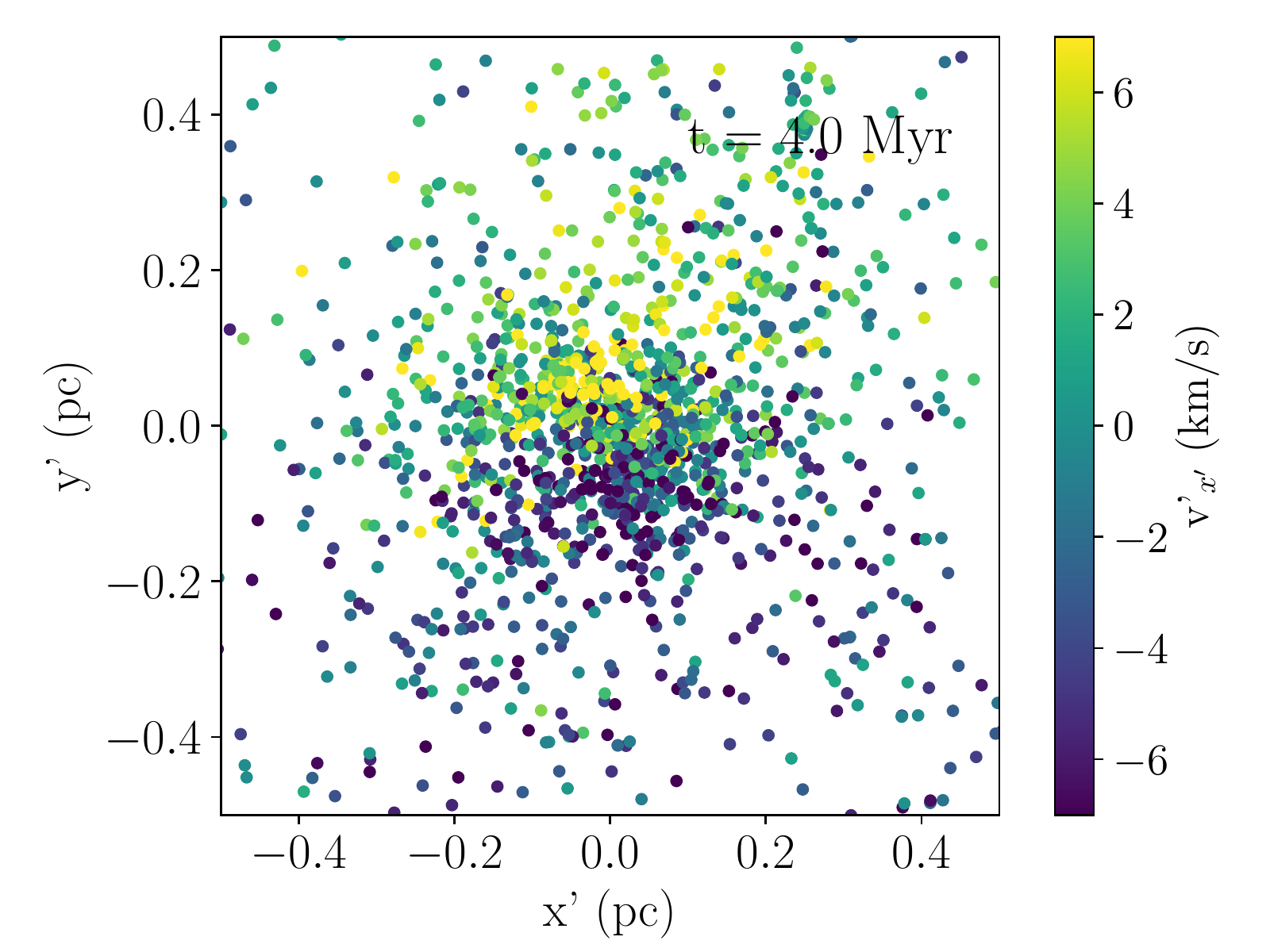}
\includegraphics[scale=0.33]{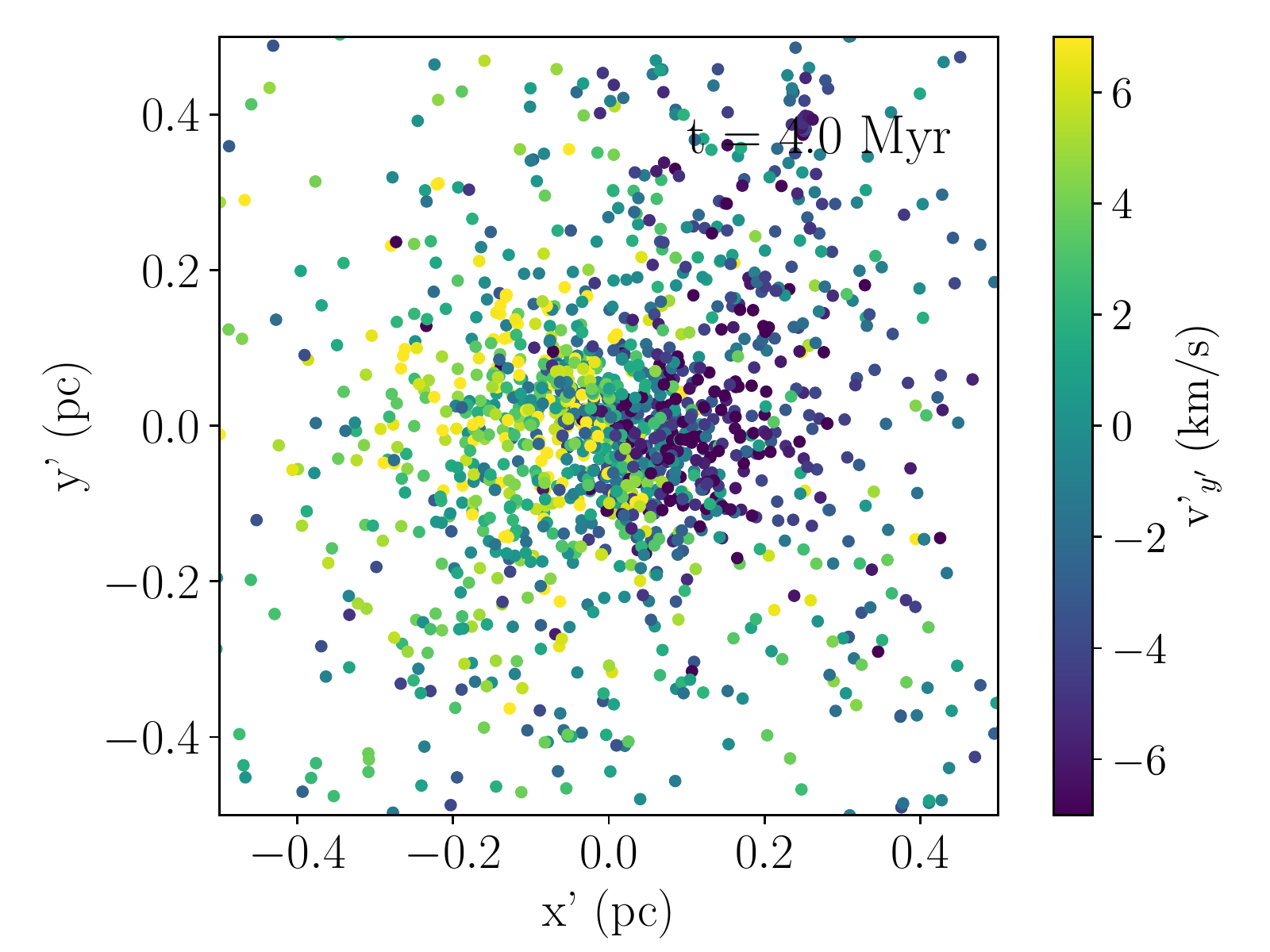}
\includegraphics[scale=0.33]{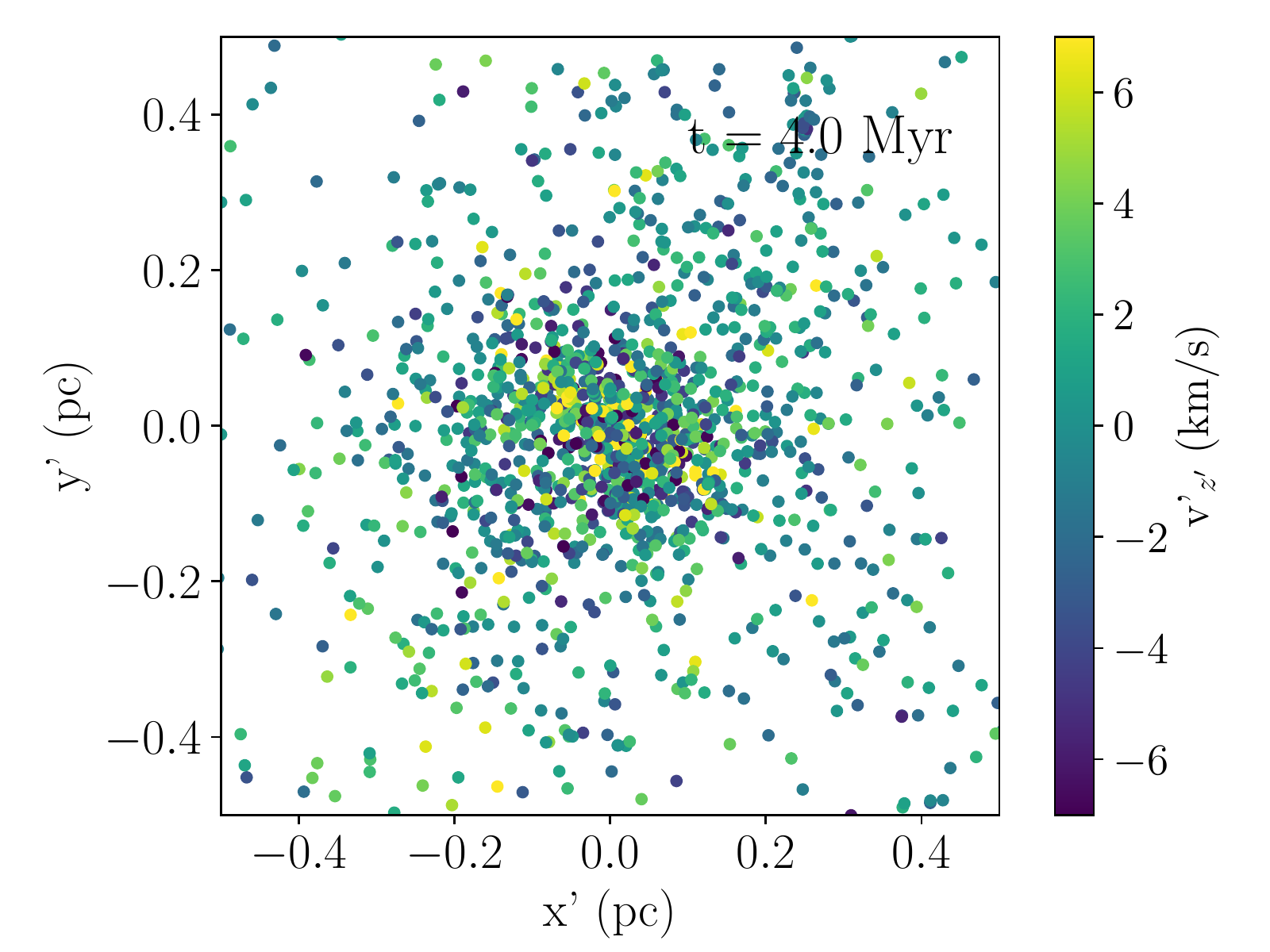}
\includegraphics[scale=0.33]{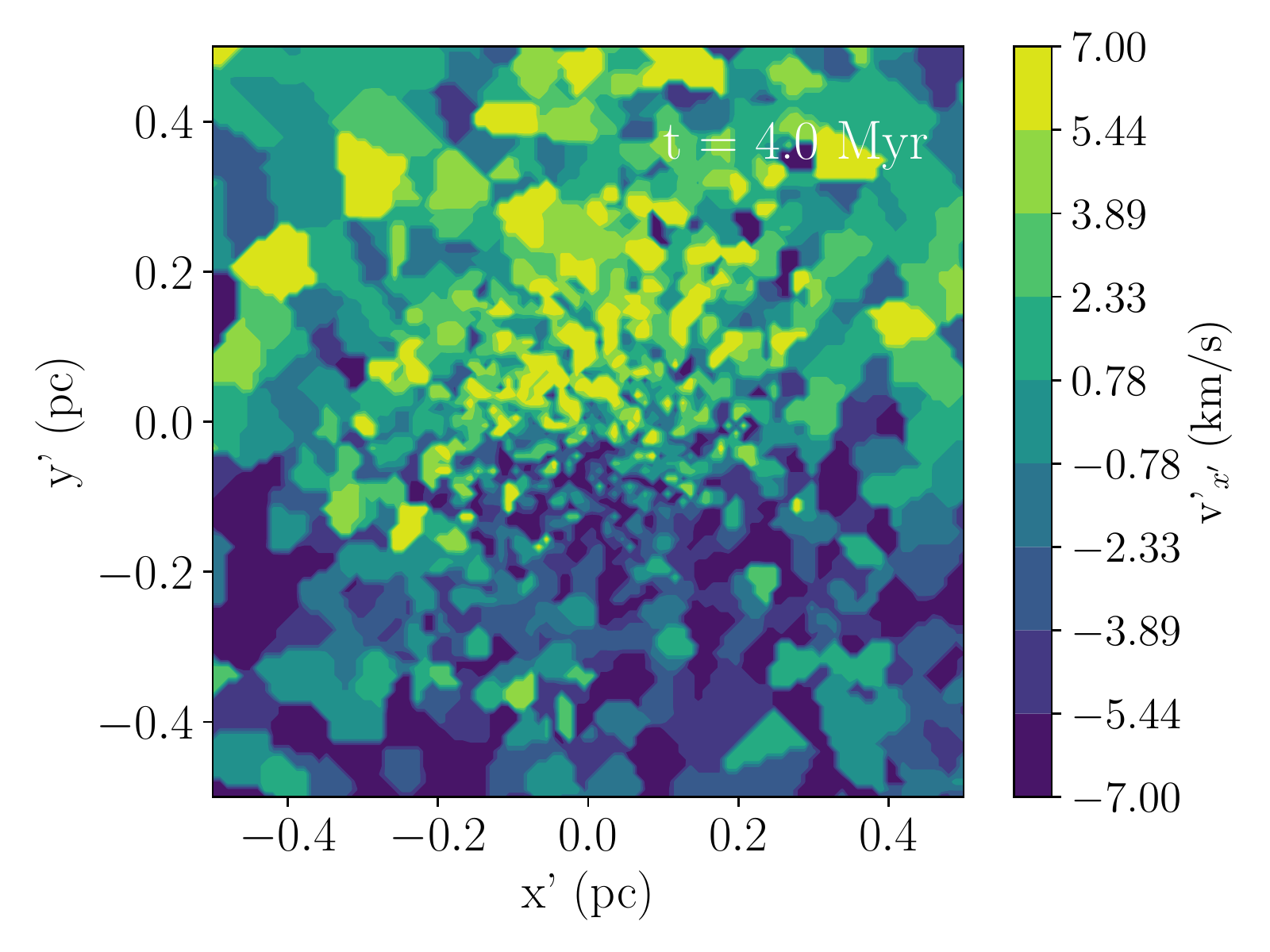}
\includegraphics[scale=0.33]{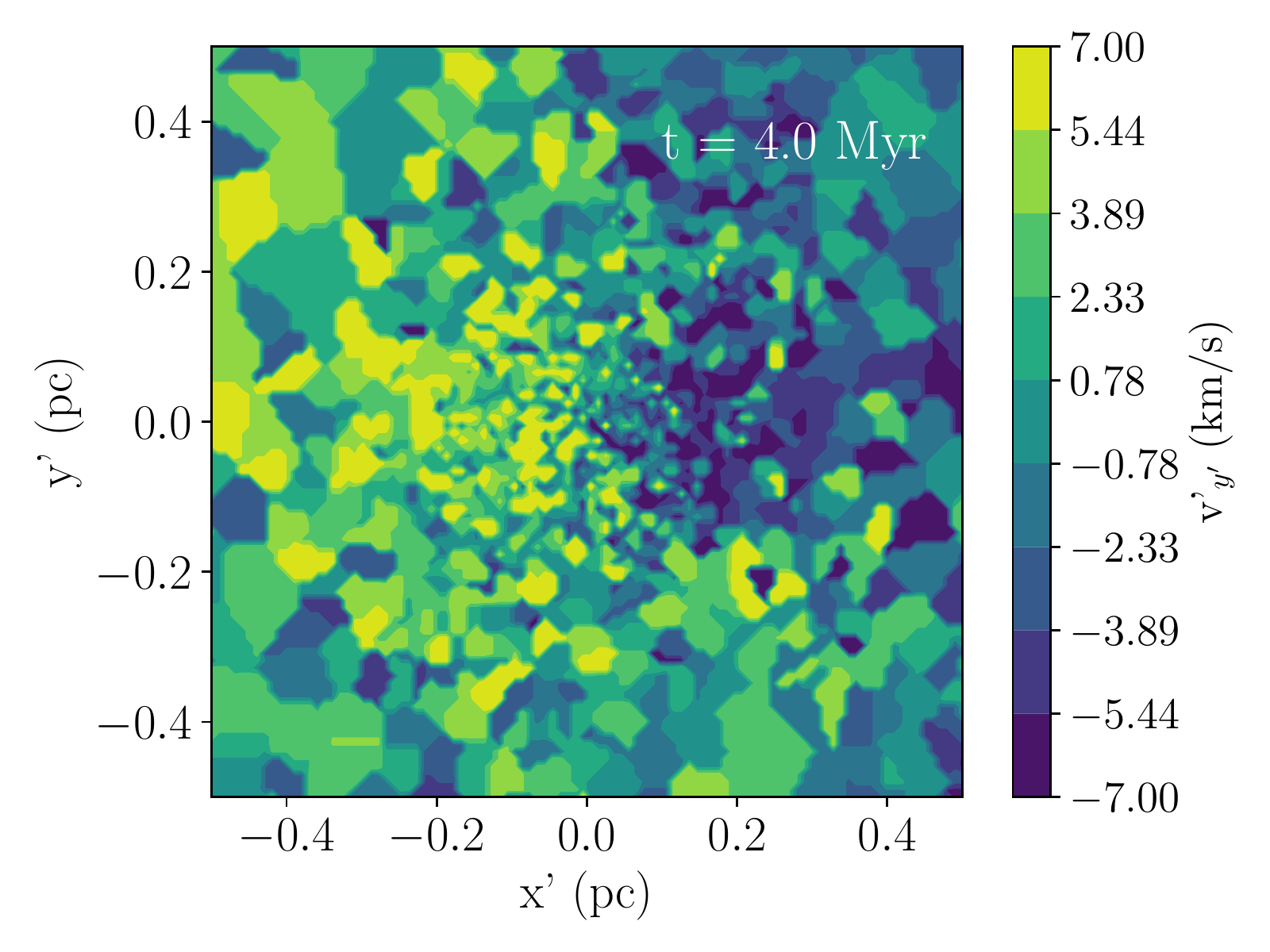}
\includegraphics[scale=0.33]{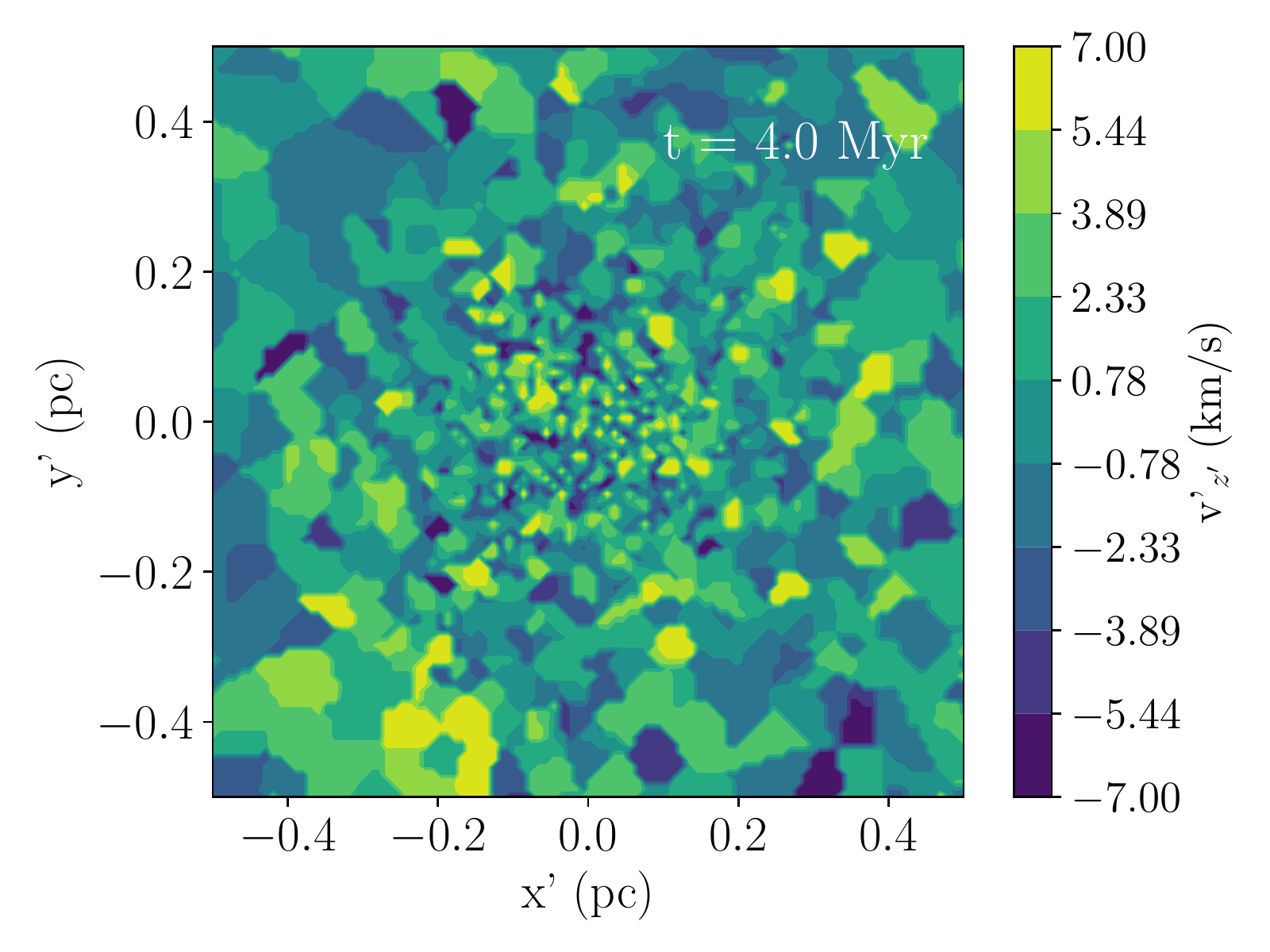}
\caption{Scatter plots (upper panels) and Voronoi tessellation maps (lower panels) of the highest angular momentum sub-cluster formed in the $M_{cl}=2\times 10^4$~M$_{\odot}$ simulation, at 4~Myr.  We show the $x'-y'$ plane, where $z'$ is the direction of the angular momentum of the sub-cluster. The colour map refers to the three components of the velocity: $v'_{x'}$ (left), $v'_{y'}$ (center) and $v'_{z'}$. The sub-cluster is the same as in Figure \ref{rot_sink}, but shown at $t_{sim}=4$ Myr.
}\label{rot_evol}
\end{center}
\end{figure*}

\begin{figure*}
\begin{center}
\includegraphics[scale=0.53]{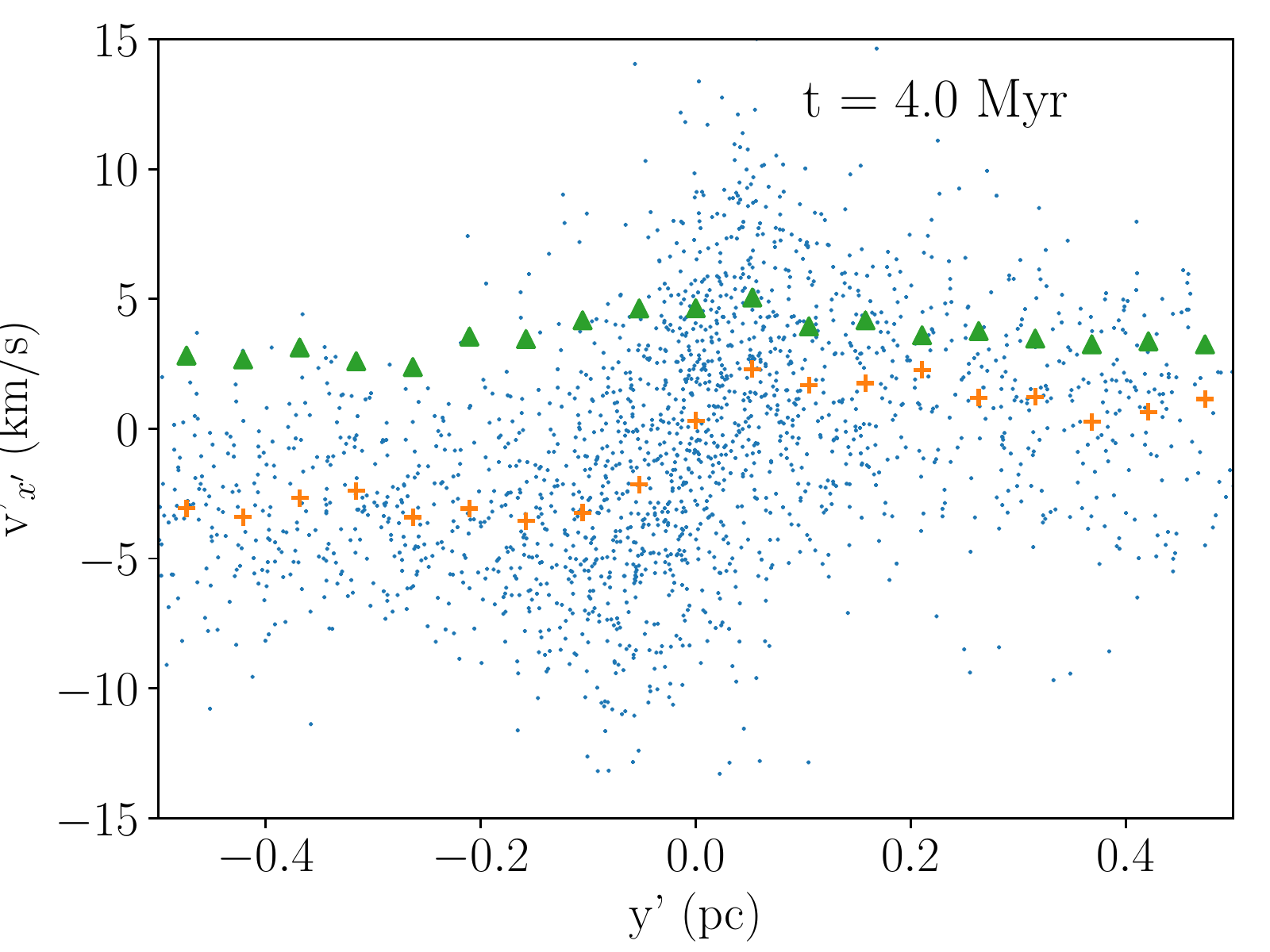}
\includegraphics[scale=0.53]{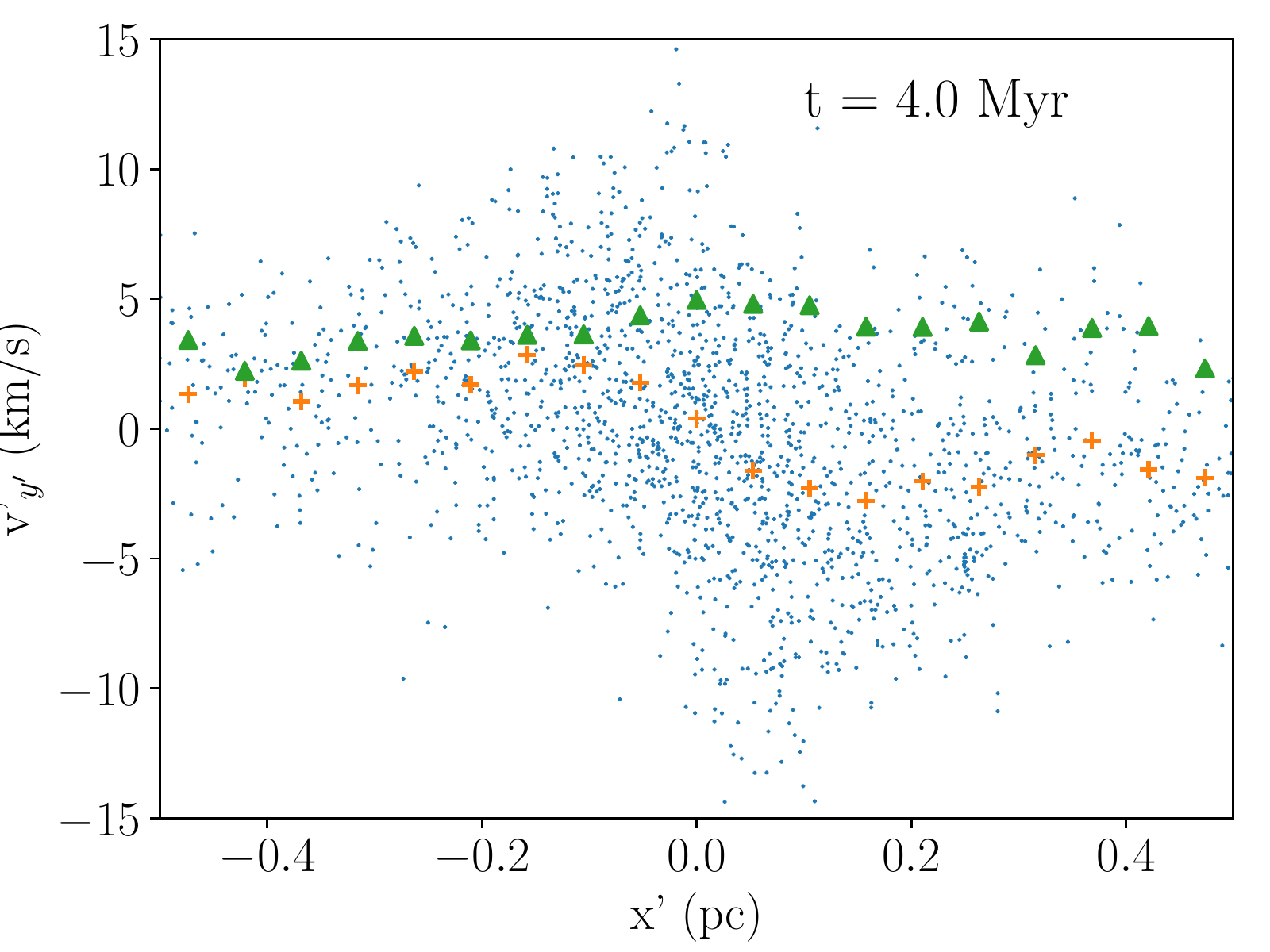}
\includegraphics[scale=0.53]{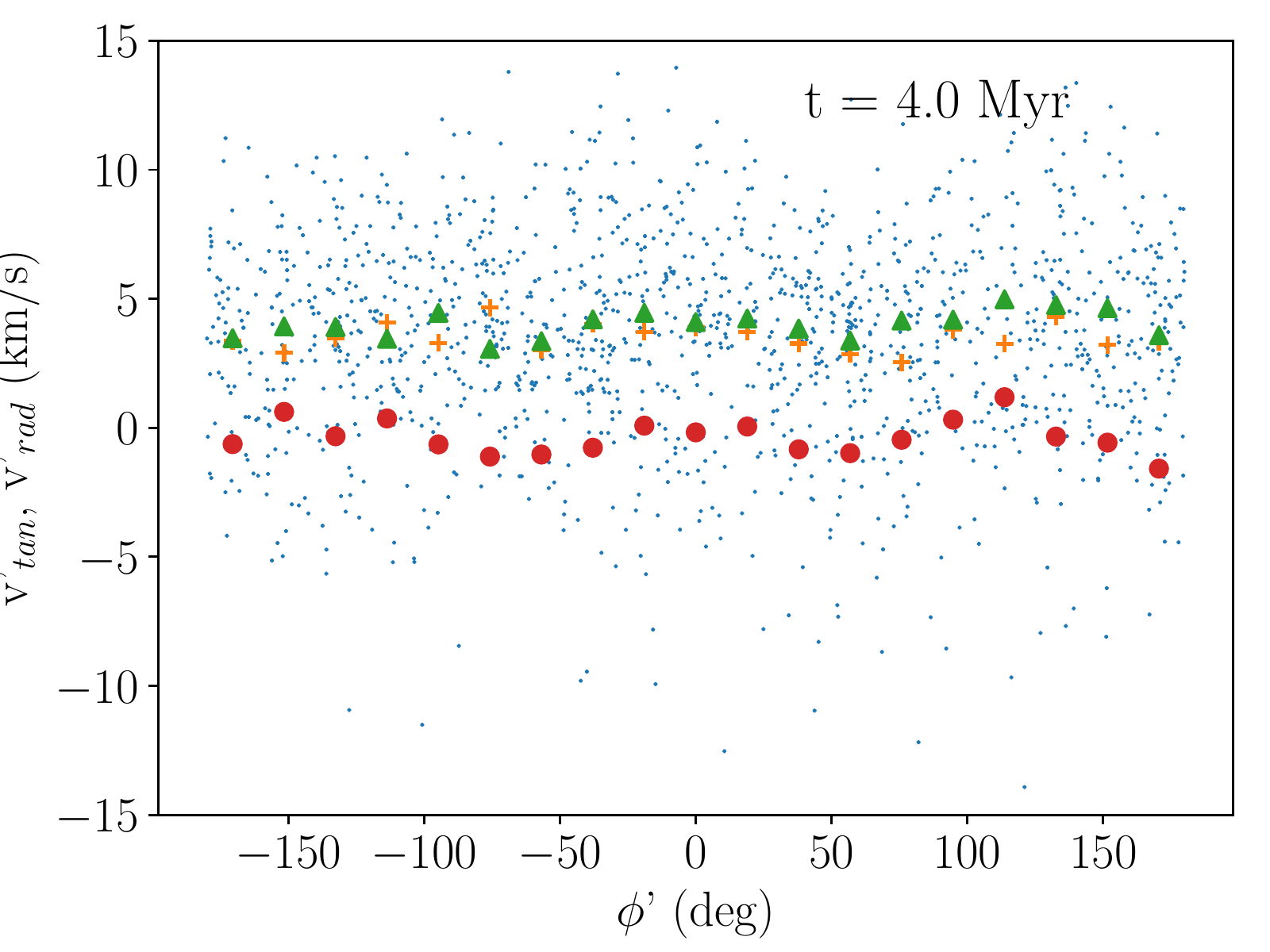}
\includegraphics[scale=0.53]{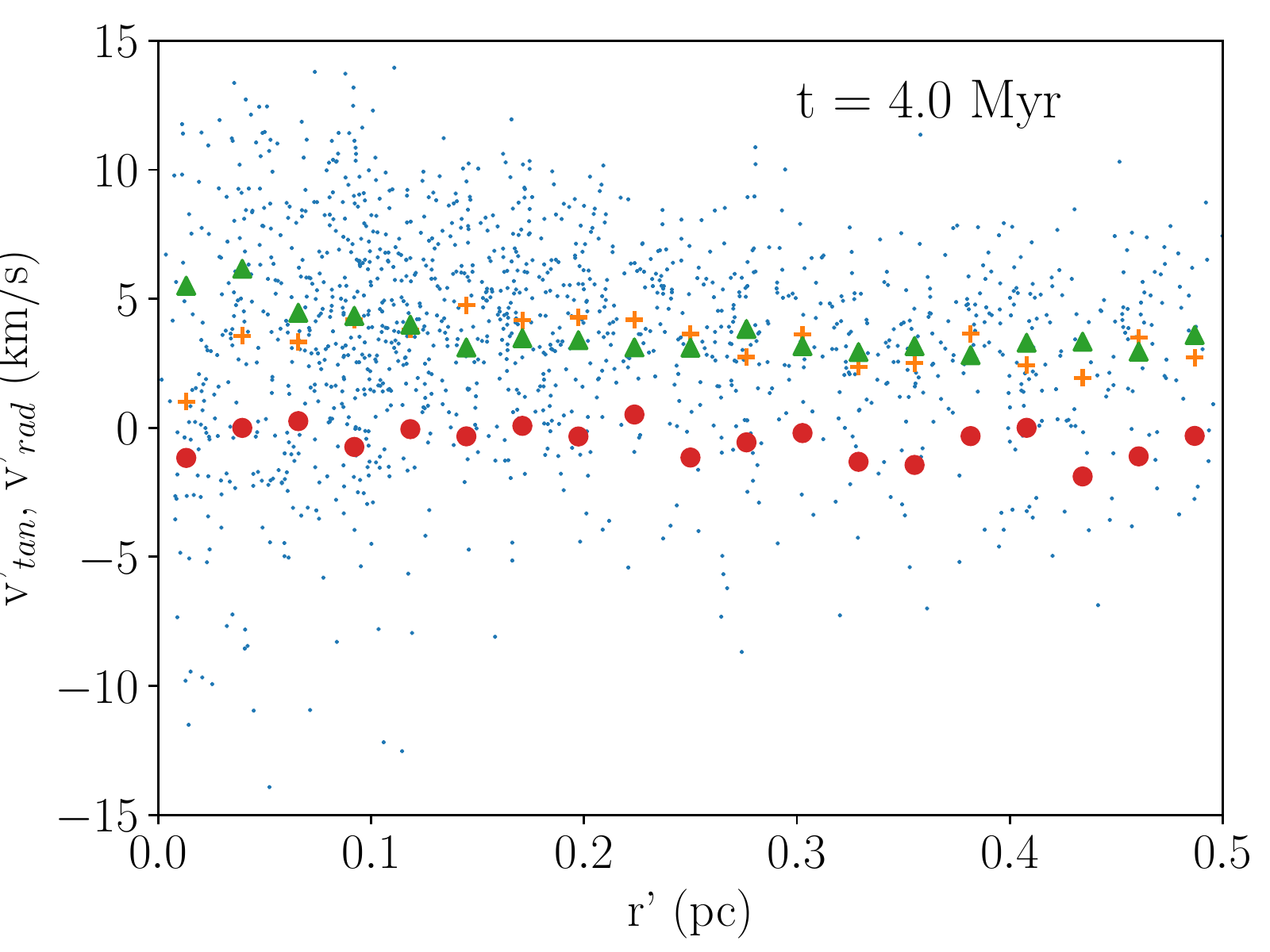}
\caption{Upper panels: Velocity profiles ($v'_{x'}$, left; $v'_{y'}$, right) of the fastest rotating sub-cluster in the $M_{cl}=2\times 10^4$~M$_{\odot}$ simulation, at 4 Myr. The blue dots show the velocity of each individual sink, the orange crosses represent the average of the velocity, while the green triangles represent the velocity dispersion at each position (see text for details). Lower panels: azimuthal (left) and radial profiles (right) for the tangential (rotational) $v'_{tan}$ and $v'_{rad}$ velocity of the same subcluster. The blue dots show the rotational velocity of each individual sink, the orange crosses represent the average rotational velocity, the green triangles represent the rotational velocity dispersion and the red circles represent the average radial velocity at each position (see text for details).
}\label{prof_vel}
\end{center}
\end{figure*}

\begin{figure*}
\begin{center}
\includegraphics[scale=0.33]{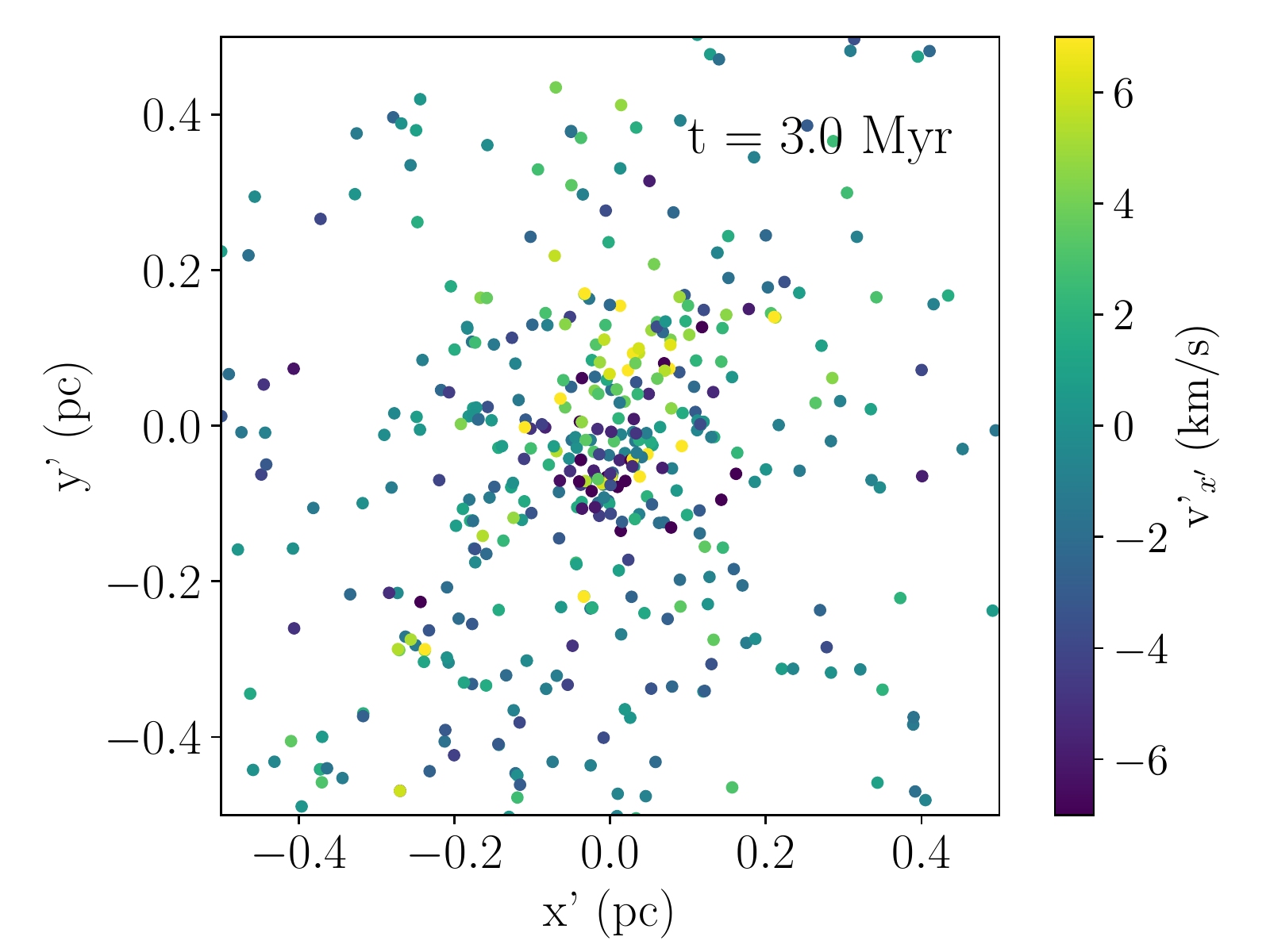}
\includegraphics[scale=0.33]{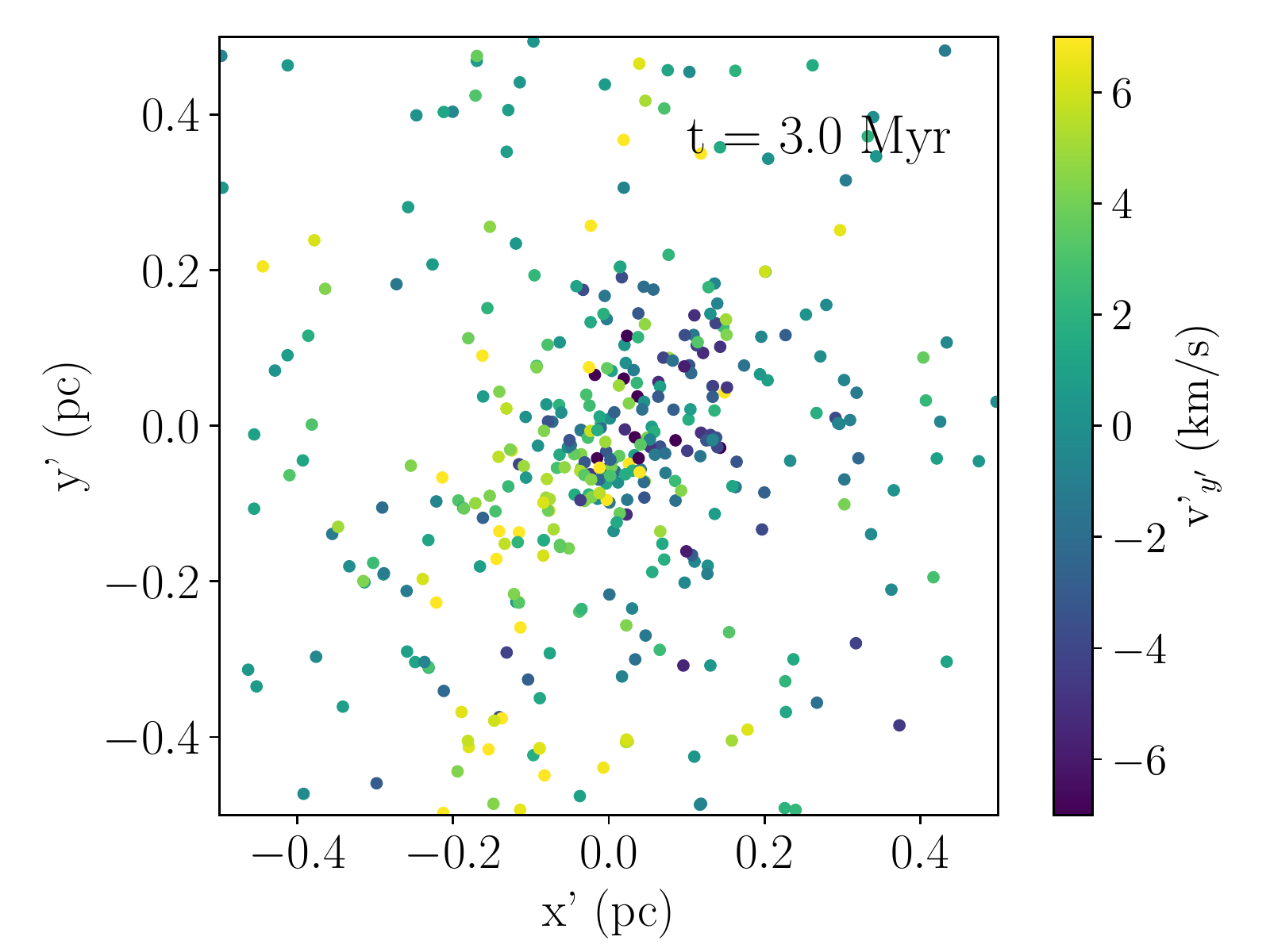}
\includegraphics[scale=0.33]{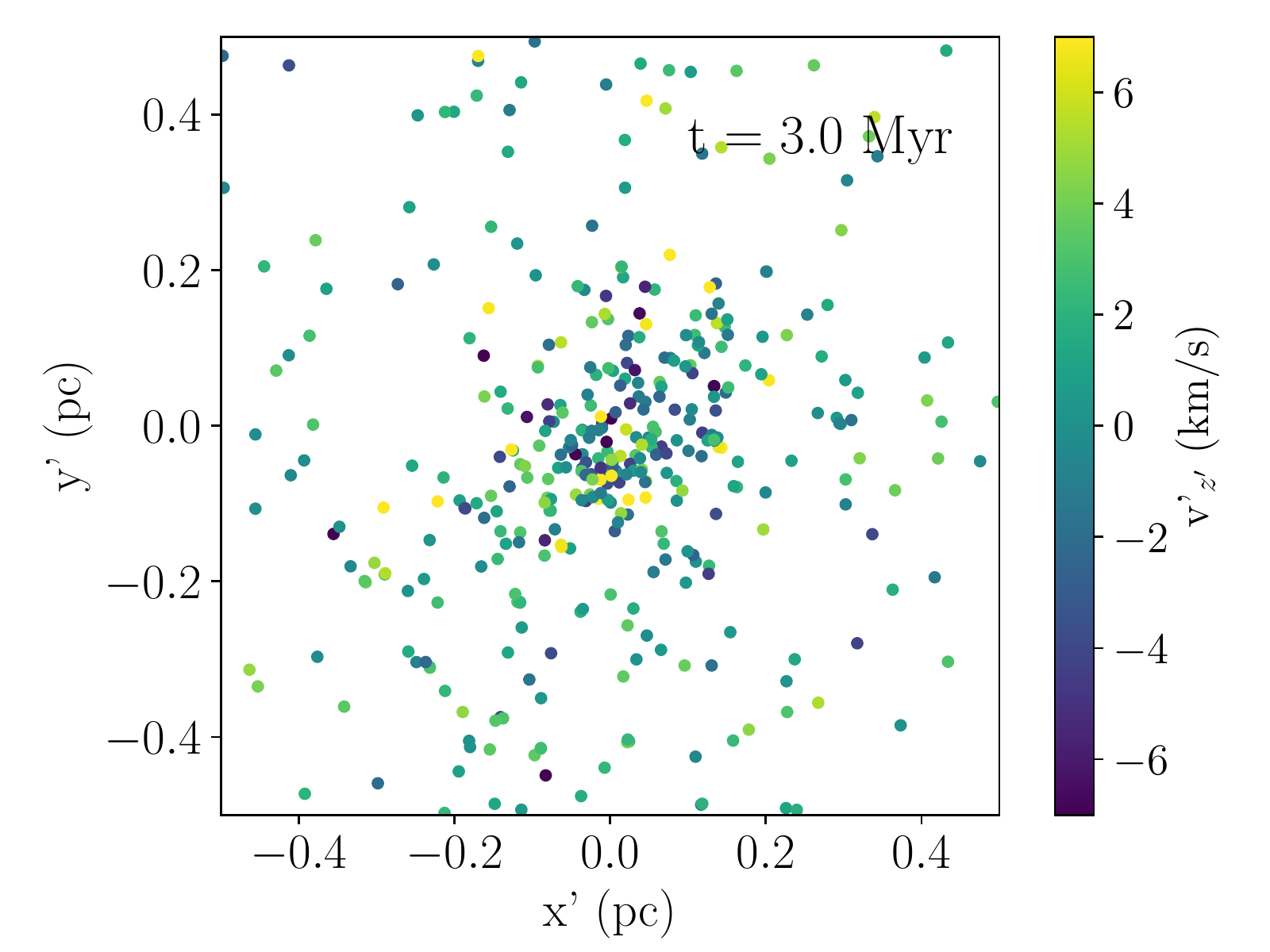}
\includegraphics[scale=0.33]{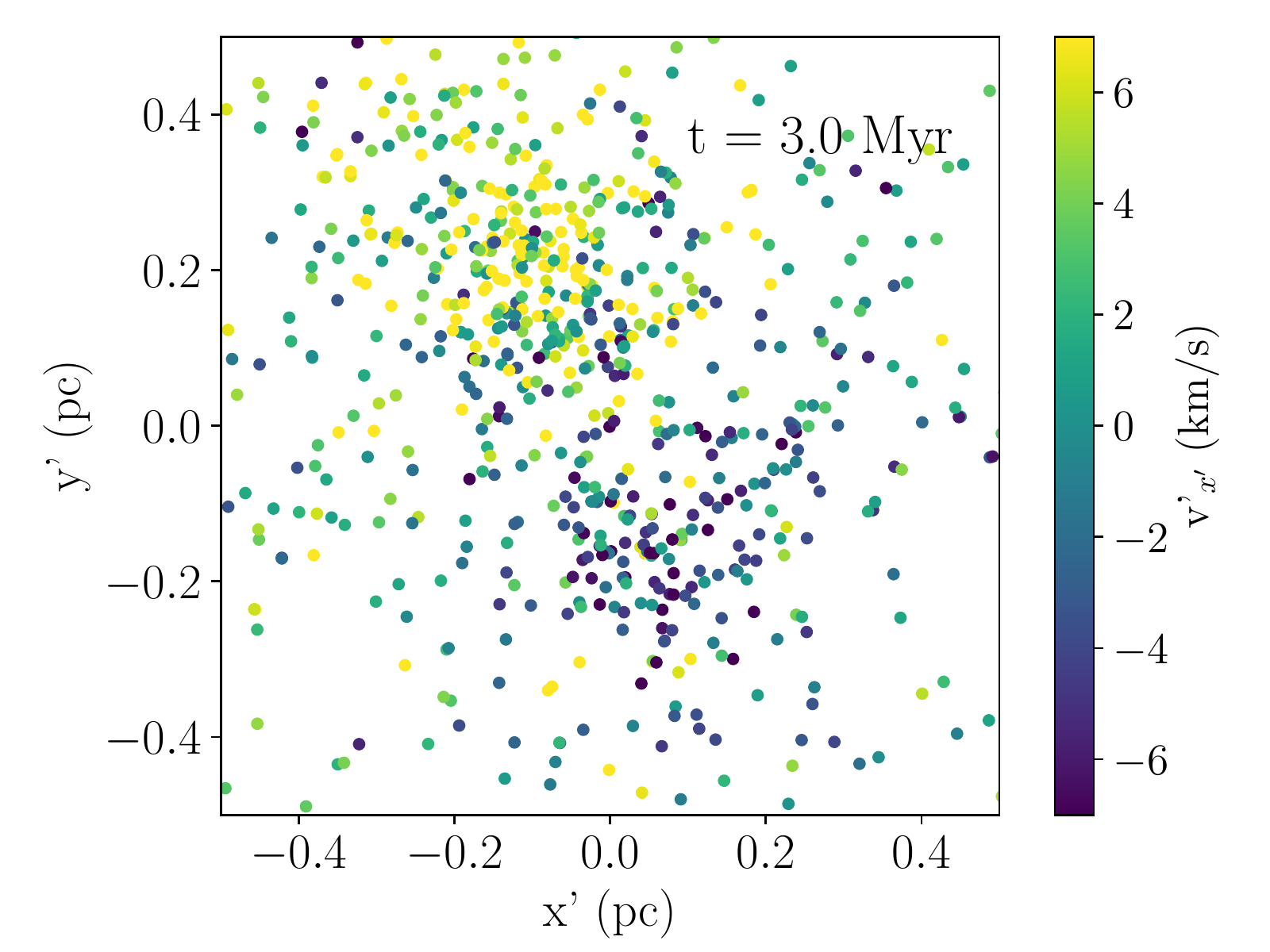}
\includegraphics[scale=0.33]{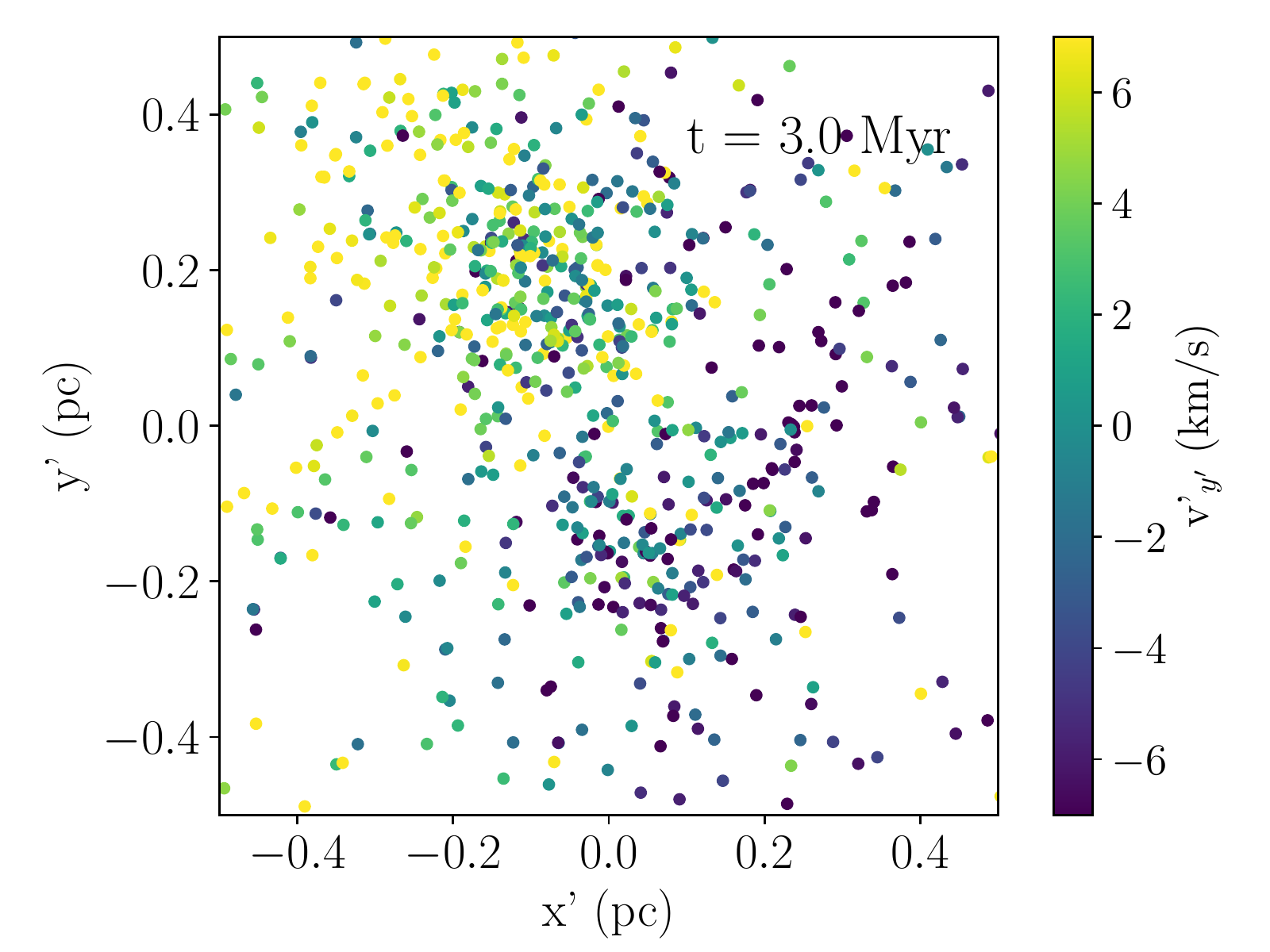}
\includegraphics[scale=0.33]{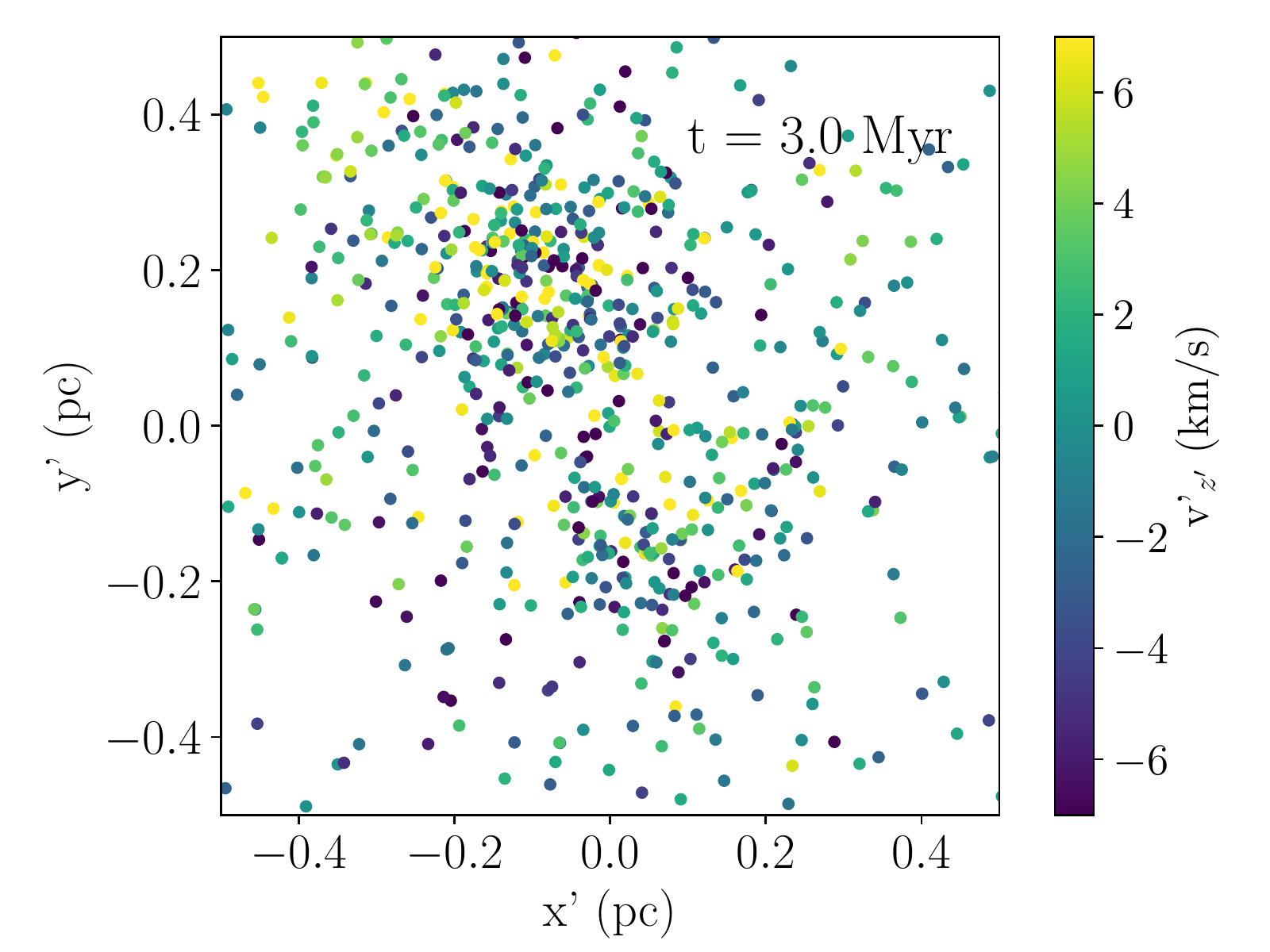}
\caption{Scatter plots for different regions in the $M_{cl}=6\times 10^4$~M$_{\odot}$ (first row) and $M_{cl}=10^5$~M$_{\odot}$ simulations (second row), at $t_{sim}=3$ Myr. The left, central and right columns show the three components of the velocity $v'_{x'}$ (left), $v'_{y'}$ (center) and $v'_{z'}$ (right) in the $x'$-$y'$ plane, where $z'$ is the direction of the angular momentum of the region.
}\label{rot_par}
\end{center}
\end{figure*}

\subsection{Fractality}\label{fracres}

All the simulated star clusters appear to be strongly substructured, compared to the relaxed distributions of older stellar systems. This is extremely important, since substructures in the cluster are usually very dense and might be the loci of strong dynamical interaction between stars. One way to quantify the degree of irregularity of an object is the so-called fractal dimension $D$. Several works in literature tried to quantify the fractal dimension, from both an observational point of view and a theoretical one. We calculated the fractal dimension of our simulated clusters in different ways.

First of all, we calculated the fractal dimension $D_B$ through a `box-counting' method. This is the inverse process of what is usually done to build fractal distributions \citep[e.g.,][]{Bate98,Cartwright04,Kupper11, Lomax11}. In this case, we constructed a series of $n=60$ 3D/2D grids embedding the star cluster, composed of cubic cells. The size $L_{k}$ of each cell, in the $k$-th grid is equal to the maximum $x,y,z$-size $S$ of the star cluster, divided by a factor ranging from 2 to 200 in a logarithmic scale. According to this definition, the fractal dimension $D_{B}$ is calculated as the slope of the curve $\log{N_{occ}}$ versus $\log L^{-1}$, where $N_{occ}$ is the number of cells occupied by at least one sink.

Figure \ref{frac_dim_B} shows the aforementioned curves for the full set of simulations, calculated for the 3D distribution of the sinks and for their distribution in the $x-y$ projection, while Table~\ref{properties} lists the values of $D_B$ obtained from a fit of those curves, calculated for $N_{occ,max}=0.1\,{}N_{sinks}$, i.e. roughly when the curves start saturating. The saturation occurs because the number of `boxes' is tending to the number of sinks at $L\lesssim 1$ pc, which means $D_B$ is mostly probing the degree of substructures on larger scales. As visible in Table~\ref{properties}, the values of $D_B$ are ranging from 1.30 to 1.56 for the 3D calculation and from 1.04 to 1.34 in the 2D case, which is on the lower end, but compatible, with the values obtained in observations \citep[see][and references therein]{Sanchez10}.
 
We then calculated the fractal dimension through a `neighbour-counting' method. This is strongly related to calculating the so-called `correlation integral' \citep[e.g.,][]{deLaFuenteMarcos06,Sanchez07}

\begin{equation}
C(r)=\frac{1}{N_s\,{}(N_s-1)}\sum_{i=1}^{N_s}n_i(r),
\end{equation}

where $N_s$ is the number of sinks and $n_i(r)$ is the number of neighbours of the \textit{i}-th sink, contained in a 3D sphere (or 2D circle) with radius $r$. Figure~\ref{frac_dim_C} shows the resulting values of $C'(r)=(N_s-1)\,{}C(r)$ as a function of $r$, both for the 3D distribution and for its 2D $x-y$ projection. Also in this case, we obtained $D_C$ as the slope of the curve $\log{C'(r)}$ versus $\log{r}$, calculated for $C'(r)_{max}=N_{sinks}/3$, i.e., roughly when the curves start saturating. In this case, the saturation occurs since for $r$ that tends to half of the size of the cluster, the corresponding sphere (circle) is basically encompassing most of the sinks. So, opposite to $D_B$, $D_C$ is mostly probing the degree of substructures on small/intermediate scales. It is also worth noting that the derivative of such curves depends on $r$. This happens because the sinks in some simulations are distributed in few (two or three) major sub-clumps. Hence, $C'(r)$ may change regime when $r$ reaches the typical size of such sub-clumps. The values of $D_C$ listed in Table~\ref{properties} are typically smaller than those of $D_B$, though still compatible with the observed values for the youngest star clusters \citep[see][and references therein]{Sanchez10}. However, this is partially the result of our simple linear fitting of curves with clearly varying slope. As also shown in \citet{Sanchez08}, $D_C$ can also strongly depend on the actual 3D distribution of the sinks (particularly, on how `flat' the distribution is, along the line of sight), especially at low values of their actual fractal dimension.

Finally, we estimated the so-called $Q$ parameter, which is likely the most popular indicator of fractality in the literature, used for the analysis of both observations and simulations \citep[e.g.,][]{Cartwright04, Schmeja06, Bastian09, Cartwright09, Sanchez10, Maschberger10, Parker12, Parker14, Parker15}. $Q$ is defined as the ratio of the mean edge length of the minimum-spanning-tree (MST) $\bar m$ and the mean inter-particle distance $\bar s$. For this calculation, $\bar s$ should be normalized to a typical size of the star cluster. We chose this length $r_{cl}$ to be half of the maximum distance between the sinks (i.e., we chose the radius of the circle, in 2D, or sphere, in 3D, encompassing the whole cluster). Following \citet{Cartwright09}, $\bar m$ was normalized by $(\pi r_{cl}^2/N_s)^{1/2}$ for a 2D calculation and by $(4\pi r_{cl}^3/(3N_s))^{1/3}$ for a 3D calculation.

In Table \ref{properties}, we listed the value of $Q$ for our set of simulations, calculated using the 3D distribution and the 2D $x-y$ projection of the sinks. We show that $Q_{2D}$ and $Q_{3D}$ are both significantly smaller than 0.8 (for the 2D calculation) and 0.7 (for the 3D calculation), which are the values expected for non-fractal distributions. The only outliers, with values $\gtrsim 0.7$, are $Q_{2d}$ for the $M_{cl}=4\times 10^4$~M$_{\odot}$ and $M_{cl}=7\times 10^4$~M$_{\odot}$ simulations. This is explained by the particular choice of their 2D projections, which made the cluster appear almost spherical and brought $Q$ to a value which is more typical of low fractality. However, the high fractality of these clusters is still revealed through the 3D calculation and the previous $D_B$ and $D_C$ indicators. 

As shown in \citet{Cartwright09}, $\bar s$-$\bar m$ plots are an even better diagnostic test of fractality. Figure~\ref{Q} can be directly compared to Figures 1 and 2 of \citet{Cartwright09}. Our values of $\bar s$ and $\bar m$ sit in a region of such plots typical of extremely high fractality, i.e., fractal dimension $\lesssim 1.6$. 

Figure \ref{D_evo} show the evolution of $D_{B,C}$ for the $M_{cl}=2\times 10^4$~M$_{\odot}$ simulation, while Figure \ref{Q_evo} show the evolution of $Q$ for the $M_{cl}=2,4,6\times 10^4$~M$_{\odot}$ simulations. Both $D_C$ and $Q$ show a clear increase with time, while $D_B$ stays almost constant during the evolution of the whole cluster. This can be explained by the fact that $D_C$ and $Q$ are mostly probing the degree of fractality on small scales, which decreases with time, due to the merger of such substructures and to their internal dynamical relaxation \citep[see also][]{Maschberger10,Parker14}. In contrast, $D_B$ mostly probes the amount of substructures on larger separations, i.e., those due to the large scale modes of the initially induced turbulence, which are more slowly affected by dynamics. As already discussed, the $M_{cl}=4\times 10^4$~M$_{\odot}$ simulation shows a big increase in $Q_{2D}$ with time, reaching values even bigger than 0.8, i.e., what is expected for no-fractality. However, as shown by the $Q_{3D}$ curve this is mostly due to the particular choice of its 2D projection.

\subsection{Rotation}\label{rotres}

We studied rotation of substructures in our simulations, performing a similar analysis as in \citet{Mapelli17}. In particular, we selected regions with the highest mass density and angular momentum of the sinks and, for these regions, we rescaled the position and velocity of the sinks to their center of mass. Finally we moved them to a new frame of reference, where $z'$ has the same direction as the total angular momentum of the selected sinks and $x'$ and $y'$ are arbitrarily oriented in the plane perpendicular to $z'$. 

Figure \ref{rot_sink} shows scatter plots and Voronoi tessellation maps for the three components of the velocity $v'_{x'}$, $v'_{y'}$ and $v'_{z'}$ in the $x'-y'$ plane, for the highest angular momentum cluster of the $M_{cl}=2\times 10^4$~M$_{\odot}$ simulation at $t_{sim}=3$ Myr. The graphs of $v'_{x'}$ and $v'_{y'}$ show a distribution of mostly positive velocity components on half of the plot and mostly negative velocity components on the other half. In contrast, the plots of $v'_{z'}$ show no clear trend. This can be interpreted as rotation of such subcluster around the $z'$ axis. 

Figure \ref{rot_evol} shows Voronoi tessellation maps of the same sub-cluster as in Figure \ref{rot_sink} but shown at a later time, at $t_{sim}=4$ Myr. The rotation signature is still present and has comparable magnitude even after 1 Myr. 

We can further confirm that the signature visible in Fig. \ref{rot_sink} and \ref{rot_evol} are actually due to rotation, by looking at Figure \ref{prof_vel}, for $t_{sim}=4$ Myr. 
In this plot we show, for each "position" bin, the average velocity components of the sinks and their dispersion $\sigma$, calculated as the standard deviation of each velocity component with respect to its average.
The upper panels of Fig. \ref{prof_vel} show profiles for $v'_{x'}(y')$ and $v'_{y'}(x')$. Such plots should be a good representation of what an observed line-of-sight velocity would look like in the plane perpendicular to the total angular momentum of the sub-cluster. The lower panels of Fig. \ref{prof_vel} show profiles for the tangential (rotational) and radial velocity components, ${v'}_{tan}$ and ${v'}_{rad}$, in the new frame of reference. The profiles are done over $\phi'$ and $r'$, where $\phi'$ is the azimuthal position and $r'$ is the distance from the center of the sub-clump, in the newly defined frame of reference.

A rotation feature is visible in all of these plots, with the average value of each velocity component associated to rotation being $\approx 3-5$~km~s$^{-1}$ and comparable to its dispersion. Such magnitudes are similar to the values obtained in both observations \citep[for the young star cluster R136 observed by][]{Henault-Brunet12} and hydrodynamical simulations \citep{Mapelli17}. Furthermore, the red circles in the lower panels show the average radial velocity in each bin, which is always close to 0, again strengthening the idea that our scatter and contour maps are actually showing a rotation feature, rather then collapse towards the center of the sub-clump.

We performed the same analysis for all our simulations, but signatures of rotation are not always easily found for substructures in all of the clouds. Figure \ref{rot_par} shows the same plots as Figure \ref{rot_sink}, but for the simulations with $M_{cl}=6, 10\times 10^4$~M$_{\odot}$.
The highest angular momentum region in the simulation with $M_{cl}=6\times 10^4$~M$_{\odot}$ is occupied by a single sub-cluster. This sub-cluster shows indications of rotation, even though the feature is more noisy, because such sub-cluster is composed of less stars than the one in Figure \ref{rot_sink}. Even if the signature is less evident in this case, the rotation velocity is roughly of the same order of magnitude as in Figure \ref{rot_sink}.

In the $M_{cl}=10^5$~M$_{\odot}$ simulation, the highest angular momentum region consists of two sub-clusters rotating around a common center of mass. These two sub-clusters are about to merge and form a single, fast rotating star cluster.

\section{Discussion}\label{discus}

We mainly focused our study on fractality and rotation, since these phenomena are mostly linked to the formation phase of young star clusters and to the interplay between stars and the gas still embedding them. Such phase is crucial to understand the assembly history of star clusters. 

We showed that all the star clusters formed in our simulations (and composed of sink particles) have highly fractal distributions, from small to large scales, particularly at early stages ($\approx 1.5 \,{}t_{ff}$). The values of different fractality indicators obtained for our clusters are consistent with observations, especially for the youngest embedded star clusters, such as Taurus, Lupus, Chamaeleon~I or the Pipe Nebula \citep[see]{Cartwright04,deLaFuenteMarcos06,Schmeja06,Sanchez10, Dib19}. 
On small scales, the degree of fractality is slowly reduced with time, as shown by the trend of $D_C$ in Figure~\ref{D_evo}. This happens because our star clusters form hierarchically (smaller substructures merge to form a major, more centrally concentrated star cluster \citealt{Schmeja06}), and because each sub-structure relaxes by efficient two-body relaxation (the two-body relaxation timescale for our main substructures is of the order of 0.5 Myr). In contrast, the degree of fractality on large scales (as shown by the trend of $D_B$ in Figure \ref{D_evo}) remains nearly constant in our simulations. This is due to the fact that the box-counting method is not capable to probe fractality on small scales (the curves in Figure \ref{frac_dim_B} saturate for L$\lesssim$ 1 pc). In more physical terms, $D_B$ is not considerably varying since the substructures on large scales are mostly the imprint of the large scale modes of the initial turbulence on the distribution of the star-forming gas, rather than of stellar dynamics. The merger of such substructures occurs on longer dynamical timescales, compared to the evolutionary times of our simulations. 

Such hierarchical assembly is crucial: for example, \citet{Fujii12} showed, by means of pure \textit{N}-body simulations, that the properties of young star clusters and open clusters are best explained when these systems are the result of mergers of smaller substructures, since the latter typically have smaller relaxation times and the merger product preserves the memory of the dynamical evolution of its constituents. Here, we show that the hierarchical assembly starts already in the embedded phase of these systems and occurs over different timescales, at different length-scales.\vspace{1cm}

Rotation can be found in substructures throughout the whole set of simulations, already at the early stages of their formation. As discussed by \citet{Mapelli17}, this is due to angular momentum conservation in the collapse of the densest gas forming the stellar substructures, as well as angular momentum transfer by torques from the gas to the already formed substructures. The magnitude of the rotation signature measured in our simulations is consistent with that found in the R136 cluster by \citet{Henault-Brunet12}.

Furthermore, rotation is visible not only in single, almost spherically symmetric substructures, but also in the rotation of different stellar sub-clumps, rotating around a common centre of mass (see Figure \ref{rot_par}). Compared to \citet{Mapelli17}, by looking at Fig. \ref{prof_vel}, we can even more strongly exclude that the rotation signature visible in Figure \ref{rot_sink}, \ref{rot_evol}, \ref{prof_vel} and \ref{rot_par} is actually due to the collapse of the substructure. This is particularly important, since an undergoing collapse along some axis might be totally mistaken for rotation \cite{Rigliaco16}.

It is worth noting that rotation persists also at later times. This is, again, a proof that stars inherit their rotation from the parent gas, until they are embedded and the gas is expelled by stellar feedback. 

Finally, we might expect that the collapse of our clouds is mostly governed by the adimensional parameter $\gamma$, defined by the ratio between the cloud-crossing timescale and its free-fall timescale. In our case, since our clouds are highly turbulent, the relevant cloud-crossing timescale can be defined as $t_{cross}\propto R_{mc}/\sigma_{turb}$, where $R_{mc}$ is the cloud radius and $\sigma_{turb}$ is the turbulence-induced velocity dispersion. The free-fall timescale is instead $t_{ff}\propto R_{mc}^{3/2}/M_{mc}^{1/2}$, where $M_{mc}$ is the cloud mass. Hence, $\gamma\propto (M_{mc}/R_{mc})^{1/2}/\sigma_{turb}\propto \sqrt{\alpha_{vir}}$. As mentioned in section \ref{meth}, the virial ratio of our simulated clouds is uniform throughout our set. So, in our simualtions the only "control" parameter is the turbulence random seed, which has a marginal impact on our results (see, e.g., Fig. \ref{frac_dim_B}, \ref{frac_dim_C}, \ref{Q}, \ref{Q_evo}), even though it leads to quite different large scale structures. Since $\gamma\propto\sqrt{\alpha_{vir}}$, we expect that similar results in terms of rotation and fractality are to be expected for any initial condition with similar virial ratios. Instead, we expect lower (higher) virial ratio star clusters to have their substructures and rotation washed out faster (slower), as shown, e.g., in the recent study by \citet{Daffern-Powell20}.

A possible caveat of our approach is in the lack of stellar feedback. Our simulations retain gas until their end and they keep on converting it into sink particles, though most of the regions of highest stellar density are already devoid of gas, at the times at which we focused our analysis. It would be interesting to check whether rotation is lost at earlier stages in more sophisticated simulations, including `gentle' pre-supernova feedback \citep[such as photoionization or stellar winds; see, e.g.,][]{VazquezSemadeni10,Dale14, Gavagnin17, Li19}. Our simulations also lack a direct-summation gravity integrator, which would allow to accurately study processes happening at the very early formation of these systems, such as the core collapse \citep{Fujii12}, the formation and evolution of binary stars \citep[e.g.,][]{Mapelli13} and binary compact objects \citep[e.g.,][]{Ziosi14, Banerjee17, Fujii17, DiCarlo19} and the runaway collision path for the formation of intermediate mass black holes \citep{Ebisuzaki01, PortegiesZwart04, Freitag06}. This could be done in the future, by either integrating such more accurate methods in the hydrodynamical simulation (this was recently attempted by \citealt{Wall19}, but it is extremely computationally challenging) or by using these simulations as initial conditions of runs with direct \textit{N}-body codes \citep[as in the case of][]{Moeckel10,Moeckel12,Parker13,Fujii16}.

\section{Summary and conclusions}\label{summary}

We studied the evolution of fractality and rotation in embedded star clusters, by means of SPH simulations of turbulent massive molecular clouds with mass ranging from $10^4$ to $10^5$ M$_\odot$. The formation of star clusters by gas fragmentation is modelled via sink particles. In our analysis, we found that all the resulting star clusters, at the early stages of their formation (1.5 $t_{cl,ff}$) have an extremely high degree of fractality ($D\approx 1.0-1.8$, $Q_{3D}\approx 0.20-0.3$). We also showed that the degree of sub-structuring slightly decreases with  time at small scales, but it stays almost constant on large scales, on timescales of the order of $1-2 \, t_{ff}$. Furthermore, we also show that these substructures are often rotating and that rotation can persist as long as the star cluster is embedded in its parent cloud, since angular momentum is continuously fed by gas converging towards the most massive stellar structures formed. The signature of rotation can be even stronger, if we start from rotating molecular clouds \citep[see][]{Li17}, which are not considered here, because we chose to adopt a conservative approach.

Fractality and rotation could have a significant impact on the evolution of the densest regions of young star clusters, by boosting the local probability of two-body encounters. This should be taken into account when studying dynamical processes believed to happen at the very early formation of these systems.
In future work, we will thus focus on studying if and how fast rotation is erased by relaxation processes, after gas removal.

\section*{Acknowledgements}

AB, MM and SR acknowledge financial support by the European Research Council for the ERC Consolidator grant DEMOBLACK, under contract no. 770017. MS acknowledges funding from the European Union’s Horizon 2020 research and innovation programme under the Marie-Sklodowska-Curie grant agreement No. 794393. We acknowledge the CINECA award HP10BQM9PE under the ISCRA initiative and the CINECA-INFN agreement, for the availability of high performance computing resources and support.




\bibliographystyle{mnras}
\bibliography{litdyn} 





\bsp	
\label{lastpage}
\end{document}